\documentclass[journal,twoside]{IEEEtran}
\usepackage{cite}
\usepackage{amsmath,amssymb,amsfonts,amsthm,mathtools}
\usepackage{algorithm}
\usepackage{algorithmic}
\usepackage{graphicx,subfig}
\usepackage{epstopdf}
\usepackage{multirow,array}
\usepackage{url}
\usepackage{dblfloatfix}
\usepackage[dvipsnames]{xcolor}

\newcommand{\prox}[2]{\operatorname{prox}_{#1}\left(#2\right)}

\newcommand{\norm}[1]{\left\lVert#1\right\rVert}
\newcommand{\abs}[1]{\left\lvert#1\right\rvert}
\newcommand{\inner}[1]{\left\langle #1 \right\rangle}

\makeatletter
\newcommand\bigcdot{\mathpalette\bigcdot@{.5}}
\newcommand\bigcdot@[2]{\mathbin{\vcenter{\hbox{\scalebox{#2}{$\m@th#1\bullet$}}}}}
\makeatother

\theoremstyle{definition}

\newtheorem{theorem}{Theorem}
\newtheorem{lemma}{Lemma}

\begin{document}

\bstctlcite{IEEE:BSTcontrol}

\title{PALMNUT: An Enhanced Proximal Alternating Linearized Minimization Algorithm with Application to Separate Regularization of Magnitude and Phase}

\author{Yunsong Liu,~\IEEEmembership{Student Member,~IEEE} and Justin P. Haldar,~\IEEEmembership{Senior Member,~IEEE}
\thanks{Y. Liu and J. P. Haldar are with the Signal and Image Processing Institute, Ming Hsieh Department of Electrical and Computer Engineering, University of Southern California, Los Angeles, CA, 90089, USA}~
\thanks{This work was supported in part by research grants NSF CCF-1350563, NIH R21-EB022951, NIH R01-MH116173, NIH R01-NS074980, and NIH R01-NS089212, as well as a USC Viterbi/Graduate School Fellowship.}}

\maketitle

\begin{abstract}
We introduce a new algorithm for complex image reconstruction with separate regularization of the image magnitude and phase.  This optimization problem is interesting in many different image reconstruction contexts, although is nonconvex and can be difficult to solve.  In this work, we first describe a novel implementation of the previous proximal alternating linearized minization (PALM) algorithm to solve this optimization problem.  We then make enhancements to PALM, leading to a new algorithm named PALMNUT that combines the PALM together with Nesterov's momentum and a novel approach that relies on uncoupled coordinatewise step sizes derived from coordinatewise Lipschitz-like bounds.  Theoretically, we establish that a version of  PALMNUT (without Nesterov's momentum) monotonically decreases the objective function, leading to guaranteed convergence in many cases of interest.  Empirical results obtained in the context of magnetic resonance imaging demonstrate that PALMNUT has computational advantages over common existing approaches like alternating minimization.  Although our focus is on the application to separate magnitude and phase regularization, we expect that the same approach may also be useful in other nonconvex optimization problems with similar objective function structure.
\end{abstract}

\section{Introduction}

Many model-based computational imaging methods assume that data acquisition can be represented as a linear system of equations, such that
\begin{equation}
\mathbf{b} = \mathbf{A}\mathbf{f} + \mathbf{n},
\end{equation}
where $\mathbf{b}$ is the length-$M$ vector of measured data samples, $\mathbf{f}$ is the length-$N$ vector of unknown image voxel values, $\mathbf{A}$ is an $M\times N$ matrix modeling the data acquisition operator, and $\mathbf{n}$ is a length-$M$ vector of additive noise perturbations.  Based on this forward model, it is relatively popular to formulate image reconstruction as a regularized linear least-squares optimization problem:
\begin{equation}
\hat{\mathbf{f}} = \arg\min_\mathbf{f} \frac{1}{2}\|\mathbf{A}\mathbf{f} - \mathbf{b}\|_2^2 + R(\mathbf{f}),\label{eq:complex}
\end{equation}
where $\|\cdot\|_2$ is the standard Euclidean ($\ell_2$) norm, and $R(\cdot)$ is a regularization functional that is designed to reduce ill-posedness by encouraging the reconstructed image to have desirable characteristics. For example, it is common to encourage the reconstructed image to be spatially-smooth by choosing $R(\cdot)$ as a penalty on the norm of the image gradient or as a penalty on the norm of the high-frequency subbands in a wavelet representation of the image.

Although the image $\mathbf{f}$ may be real-valued  in many scenarios (i.e., $\mathbf{f} \in \mathbb{R}^N$), there are several imaging modalities (including certain kinds of optical imaging, ultrasound imaging, magnetic resonance imaging (MRI), synthetic aperture radar, etc.) for which the image of interest is most naturally modeled as complex-valued (i.e., $\mathbf{f} \in \mathbb{C}^N$).  In such cases, the image magnitude and image phase often represent manifestations of different physical phenomena, and have distinct spatial characteristics from one another that are not easily captured by a single unified regularization penalty.  This  has led several authors, working on image reconstruction across a variety of different modalities and application domains, to consider optimization problems in which the magnitude and phase of the image are regularized separately \cite{cetin2001,fessler2004,tuysuzoglu2012,JustinDecovSLIM2011,FesslerSepMagPhs2012,guven2016, ZibettiSepMagPhs2017,LustigPhaseCycling2018,moradikia2018,haldar2019}.  

While these approaches were developed independently in many cases, the optimization problems  can largely be unified by the following general formulation:
\begin{equation} \label{eq:GeneralReconModel}
\{\hat{\mathbf{m}},\hat{\mathbf{p}}\} = \arg\min_{\substack{\mathbf{m} \in \mathbb{R}^N \\\mathbf{p} \in \mathbb{R}^N}} \tilde{J}(\mathbf{m,p}) + R_1(\mathbf{m}) + \tilde{R}_2(\mathbf{p}),
\end{equation}
where 
\begin{equation}\label{eq:F}
\tilde{J}(\mathbf{m,p}) \triangleq \frac{1}{2}\|\mathbf{A}\left(\mathbf{m}\odot e^{i\mathbf{p}}\right) - \mathbf{b}\|_2^2.
\end{equation}
In this expression, $\mathbf{m}$ and $\mathbf{p}$ respectively represent the magnitude\footnote{Note that this formulation allows the ``magnitude'' vector $\mathbf{m}$ to have negative entries, which can avoid unnecessary phase discontinuities in $\mathbf{p}$  \cite{ FesslerSepMagPhs2012}.} and phase of the complex image $\mathbf{f}$ such that $\mathbf{f} = \mathbf{m}\odot e^{i\mathbf{p}}$, and $R_1(\cdot)$ and $\tilde{R}_2(\cdot)$ respectively represent separate regularization penalties for the magnitude and phase.  We have used the notation $\odot$ to represent the Hadamard product (elementwise multiplication) of two vectors, have used the notation $e^\mathbf{z}$ to denote the  elementwise exponentiation of the vector $\mathbf{z}$, and have used $i$ to denote the unit imaginary number ($i=\sqrt{-1}$).  

On one hand, Eq.~\eqref{eq:GeneralReconModel} is desirable because it allows independent customization of the regularization penalties applied to the magnitude and phase components of the image, which has previously been shown to be beneficial  \cite{cetin2001,fessler2004,tuysuzoglu2012,JustinDecovSLIM2011,FesslerSepMagPhs2012, guven2016,ZibettiSepMagPhs2017,LustigPhaseCycling2018,moradikia2018,haldar2019}.  On the other hand, this  formulation can be substantially more difficult to solve than Eq.~\eqref{eq:complex}.  For example, while Eq.~\eqref{eq:complex} is convex if $R(\cdot)$ is chosen to be convex (and therefore can be globally optimized from arbitrary initializations using standard optimization methods!), the formulation in Eq.~\eqref{eq:GeneralReconModel} is generally non-convex and is thus more challenging to optimize.  As a result, many authors have decided to make use of a simple alternating minimization (AM) strategy that alternates between optimizing $\mathbf{m}$ for a fixed value of $\mathbf{p}$ and optimizing $\mathbf{p}$ for a fixed value of $\mathbf{m}$ \cite{fessler2004,JustinDecovSLIM2011,FesslerSepMagPhs2012,ZibettiSepMagPhs2017, LustigPhaseCycling2018, haldar2019}.  However, while AM can successfully decrease the objective function, it is not necessarily computationally efficient.

In this work, we propose and evaluate a new algorithm named PALMNUT to solve Eq.~\eqref{eq:GeneralReconModel}.  This algorithm is based on combining the proximal alternating linearized minimization (PALM) algorithm \cite{BoltePALM2014} with Nesterov's momentum \cite{NesterovBook2004} and a novel approach that uses uncoupled coordinatewise step sizes derived from coordinatewise Lipschitz-like bounds.  PALMNUT is evaluated in the context of several different MRI-related inverse problems, where it is demonstrated to  outperform popular existing methods.  A preliminary account of portions of this work was previously presented in a recent conference \cite{liuISMRM2019}.

\section{Background}\label{sec:back}
The following subsections review the AM approach (one of the dominant existing algorithms for solving Eq.~\eqref{eq:GeneralReconModel}) as well as the related PALM algorithm (which serves as the foundation for our proposed PALMNUT approach).  

Since both AM and PALM are generic algorithms that are broadly applicable beyond just the magnitude and phase problem from Eq.~\eqref{eq:GeneralReconModel}, we will start by introducing both of these algorithms in a more general setting before specializing to our problem of interest.  Specifically, we will consider the generic optimization problem of the form 
\begin{equation}\label{eq:AMcost}
\{\hat{\mathbf{x}},\hat{\mathbf{y}}\} = \arg\min_{\substack{\mathbf{x} \in \mathbb{R}^{N_{1}} \\ \mathbf{y} \in \mathbb{R}^{N_{2}}}} \Psi(\mathbf{x,y}),
\end{equation}
where $\mathbf{x} \in \mathbb{R}^{N_{1}}$ and $\mathbf{y} \in \mathbb{R}^{N_{2}}$ are optimization variables, and the objective function $\Psi(\cdot,\cdot)$ can be decomposed as 
\begin{equation}\label{eq:PALMcost}
	\Psi(\mathbf{x,y}) = H(\mathbf{x},\mathbf{y}) + F(\mathbf{x}) + G(\mathbf{y})
\end{equation}
for some real-valued or extended-real-valued scalar functions $H(\cdot,\cdot)$, $F(\cdot)$, and $G(\cdot)$.

\subsection{Alternating Minimization}
Given the problem setting described above, the AM algorithm proceeds according to Alg.~\ref{alg:AM}.

\begin{algorithm}
	\caption{Alternating Minimization for Eq.~\eqref{eq:AMcost}}
	\label{alg:AM}
	\begin{algorithmic}
		\REQUIRE Set $k=1$ and initialize $\hat{\mathbf{x}}_0$ and $\hat{\mathbf{y}}_0$.
		\WHILE {not converge}
		\vspace{0.05in}
		\STATE $\displaystyle \hat{\mathbf{x}}_{k} = \arg\hspace{-0.5em}\min_{\mathbf{x} \in \mathbb{R}^{N_1}} \Psi(\mathbf{x},\hat{\mathbf{y}}_{k-1})$\\
		\hspace{2.4em}$\displaystyle = \arg\hspace{-0.5em}\min_{\mathbf{x} \in \mathbb{R}^{N_1}} H(\mathbf{x},\hat{\mathbf{y}}_{k-1})+F(\mathbf{x})$ \\
		\STATE $\displaystyle \hat{\mathbf{y}}_{k} = \arg\hspace{-0.5em}\min_{\mathbf{y} \in \mathbb{R}^{N_2}} \Psi(\hat{\mathbf{x}}_{k},\mathbf{y})$\\
		\hspace{2.4em}$\displaystyle = \arg\hspace{-0.5em}\min_{\mathbf{y} \in \mathbb{R}^{N_2}} H(\hat{\mathbf{x}}_{k},\mathbf{y})+G(\mathbf{y})$ \\
		\STATE $k \gets k + 1$
		\ENDWHILE
	\end{algorithmic}
\end{algorithm}

As can be seen, AM alternates between optimizing the estimate of $\mathbf{x}$ for a fixed estimated value of $\mathbf{y}$ and optimizing the estimate of $\mathbf{y}$ for a fixed estimated value of $\mathbf{x}$. This can be viewed as a  block coordinate descent algorithm, and thus has well-studied theoretical characteristics \cite{wright2015}. The algorithm leads to monotonic decrease of the objective function value, and therefore (by the monotone convergence theorem) convergence of the objective function value if $\Psi(\mathbf{x},\mathbf{y})$ is bounded from below.   In practice, it is not necessary to solve the optimization subproblems for $\hat{\mathbf{x}}_{k}$ and $\hat{\mathbf{y}}_{k}$ exactly in each step, as the overall objective function will still decrease monotonically as long as the subproblem objective function values always decrease in each iteration. 

In the case of optimizing Eq.~\eqref{eq:GeneralReconModel} with AM  \cite{fessler2004,JustinDecovSLIM2011,FesslerSepMagPhs2012,ZibettiSepMagPhs2017, LustigPhaseCycling2018, haldar2019}, one should associate $\mathbf{x}$ with $\mathbf{m}$, $\mathbf{y}$ with $\mathbf{p}$, $H(\cdot,\cdot)$ with $\tilde{J}(\cdot,\cdot)$, $F(\cdot)$ with $R_1(\cdot)$, and $G(\cdot)$ with $\tilde{R}_2(\cdot)$. 

When defining $\tilde{J}(\mathbf{m},\mathbf{p})$ as in Eq.~\eqref{eq:F}, the magnitude subproblem has the form of a classical regularized linear least-squares problem.  As such, there exist many different algorithms to solve this kind of problem efficiently.  However, the phase subproblem is more complicated, as it is generally nonlinear and nonconvex.  Existing methods have frequently relied on either the nonlinear conjugate gradient (NCG) algorithm \cite{fessler2004,JustinDecovSLIM2011,FesslerSepMagPhs2012, ZibettiSepMagPhs2017,haldar2019} or a phase cycling heuristic \cite{LustigPhaseCycling2018} to address this subproblem. 

\subsection{Proximal Alternating Linearized Minimization (PALM)}
The PALM algorithm considers the same setup described previously, but with the additional assumptions that $H(\cdot,\cdot)$ is a smooth real-valued function and $F(\cdot)$ and $G(\cdot)$ are  proper and lower semicontinuous (but potentially nonsmooth) extended-real-valued functions. None of these functions are required to be convex.   

The structure of PALM is strongly motivated by the AM algorithm, and similar to AM, the PALM algorithm for Eq.~\eqref{eq:AMcost} proceeds by updating the $\mathbf{x}$ and $\mathbf{y}$ variables in alternation.  However, rather than directly using the AM update formulas, PALM first applies proximal linearization of $H$ to each subproblem prior to computing each update step.  Specifically, PALM uses
\begin{equation}
\begin{split}\label{eq:prox_phs}
	\hat{\mathbf{x}}_{k} = \arg\hspace{-0.5em}\min_{\mathbf{x} \in \mathbb{R}^N_{1}} & H(\hat{\mathbf{x}}_{k-1},\hat{\mathbf{y}}_{k-1}) \\ &+ \inner{\mathbf{x} - \hat{\mathbf{x}}_{k-1}, \nabla_\mathbf{x} H(\hat{\mathbf{x}}_{k-1},\hat{\mathbf{y}}_{k-1})} \\
	&+ \frac{c_k}{2} \norm{\mathbf{x} - \hat{\mathbf{x}}_{k-1}}_2^2 +  F(\mathbf{x}) + G(\hat{\mathbf{y}}_{k-1})
	\end{split}
\end{equation}
and
\begin{equation}\begin{split}\label{eq:prox_mag}
	\hat{\mathbf{y}}_{k} = \arg\hspace{-0.5em}\min_{\mathbf{y} \in \mathbb{R}^{N_{2}}} &H(\hat{\mathbf{x}}_{k},\hat{\mathbf{y}}_{k-1}) \\ &+ \inner{\mathbf{y} - \hat{\mathbf{y}}_{k-1}, \nabla_\mathbf{y} H(\hat{\mathbf{x}}_{k},\hat{\mathbf{y}}_{k-1})}  \\
	&+ \frac{d_k}{2} \|\mathbf{y} - \hat{\mathbf{y}}_{k-1}\|_2^2 + F(\hat{\mathbf{x}}_{k}) + G(\mathbf{y}),
\end{split}
\end{equation}
where $c_k$ and $d_k$ are real-valued positive scalars,  $\nabla_\mathbf{x}$ represents the gradient with respect to $\mathbf{x}$, and $\inner{\cdot,\cdot}$ denotes the standard dot product.

One way to justify Eqs.~\eqref{eq:prox_phs} and \eqref{eq:prox_mag} is to invoke the majorize-minimize algorithmic framework \cite{hunter2004}.    Specifically, for any real-valued objective function $Q(\mathbf{x})$ with $\mathbf{x}\in\mathbb{R}^N$, we say that the function $S_{k}(\mathbf{x})$ is a ``majorant'' of $Q(\mathbf{x})$ at the point $\hat{\mathbf{x}}_{k-1}$ if it satisfies two conditions: (i) $S_{k}(\hat{\mathbf{x}}_{k-1}) = Q(\hat{\mathbf{x}}_{k-1})$; and (ii) $Q(\mathbf{x}) \leq S_{k}(\mathbf{x})$ for $\forall \mathbf{x} \in \mathbb{R}^N$.  An important feature of majorant functions is that if $S_{k}(\hat{\mathbf{x}}_{k}) <S_{k}(\hat{\mathbf{x}}_{k-1})$, then it is guaranteed that $Q(\hat{\mathbf{x}}_{k}) < Q(\hat{\mathbf{x}}_{k-1})$.  This fact allows the majorant $S_{k}(\mathbf{x})$ to be used as a surrogate for the original objective function $Q(\mathbf{x})$, since descending on $S_{k}(\mathbf{x})$ guarantees descent on $Q(\mathbf{x})$. The use of a surrogate can be beneficial whenever the surrogate is easier to optimize than the original function \cite{hunter2004}.  

With these concepts defined, we return to Eqs.~\eqref{eq:prox_phs} and \eqref{eq:prox_mag}.  Assume that for a fixed value of $\mathbf{y}$, $\nabla_\mathbf{x} H(\mathbf{x},{\mathbf{y}})$ is globally Lipschitz in $\mathbf{x}$ such that
\begin{equation} \label{eq:PALM_Lipx}
\norm{ \nabla_\mathbf{x} H(\mathbf{x}_1,{\mathbf{y}}) - \nabla_\mathbf{x} H(\mathbf{x}_2,{\mathbf{y}}) }_2 \leq L_1(\mathbf{y}) \norm{ \mathbf{x}_1 - \mathbf{x}_2 }_2
\end{equation}
for $\forall \mathbf{x}_1,\mathbf{x}_2 \in \mathbb{R}^{N_1}$, where $L_1(\mathbf{y})$ is the corresponding Lipschitz constant.  Similarly, assume that for a fixed value of $\mathbf{x}$, $\nabla_\mathbf{y} H(\mathbf{x},{\mathbf{y}})$ is globally Lipschitz in $\mathbf{y}$ such that
\begin{equation} \label{eq:PALM_Lipy}
	\norm{ \nabla_\mathbf{y} H(\mathbf{x},\mathbf{y}_1) - \nabla_\mathbf{y} H(\mathbf{x},\mathbf{y}_2) }_2 \leq L_2(\mathbf{x}) \norm{ \mathbf{y}_1 - \mathbf{y}_2 }_2
\end{equation}	
for $\forall \mathbf{y}_1,\mathbf{y}_2 \in \mathbb{R}^{N_2}$, where $L_2(\mathbf{x})$ is the corresponding Lipschitz constant.  Under these assumptions,  it can be shown that the objective function in Eq.~\eqref{eq:prox_phs} will be a majorant of $\Psi(\mathbf{x},\hat{\mathbf{y}}_{k-1})$ at the point $\mathbf{x} = \hat{\mathbf{x}}_{k-1}$ whenever $c_k \geq L_1(\hat{\mathbf{y}}_{k-1})$, and the objective function in Eq.~\eqref{eq:prox_mag} will be a majorant of $\Psi(\hat{\mathbf{x}}_{k},\mathbf{y})$ at the point $\mathbf{y} = \hat{\mathbf{y}}_{k-1}$ whenever $d_k \geq L_2(\hat{\mathbf{x}}_{k})$ \cite{BoltePALM2014}.   This means that if $c_k$ and $d_k$ are chosen large enough, then Eqs.~\eqref{eq:prox_phs} and \eqref{eq:prox_mag} can be interpreted as majorize-minimize steps.  This will guarantee monotonic decrease of the objective function value, and therefore convergence of the objective function value will be guaranteed for PALM if $\Psi(\mathbf{x},\mathbf{y})$ is bounded from below  \cite{hunter2004}.  Under some additional mild conditions, the iterates of PALM are guaranteed to converge to a critical point of $\Psi(\mathbf{x},\mathbf{y})$ \cite{BoltePALM2014}. 

The solutions to the optimization problems in Eqs.~\eqref{eq:prox_phs} and \eqref{eq:prox_mag} can be written in a more concise ``proximal operator'' form \cite{BoltePALM2014}.  Specifically, Eq.~\eqref{eq:prox_phs} is equivalent to the ``proximal operator'' 
\begin{equation} \label{eq:PALM_xUpdate}
\begin{split}
\hat{\mathbf{x}}_{k} &= \arg\hspace{-0.5em}\min_{\mathbf{x} \in \mathbb{R}^{N_1}} F(\mathbf{x}) + \frac{c_k}{2} \norm{ \mathbf{x} - \mathbf{w}_k }_2^2\\
&\triangleq \prox{F}{\mathbf{w}_k,c_k}
\end{split}
\end{equation}
with  
\begin{equation}\label{eq:w}
\mathbf{w}_k = \hat{\mathbf{x}}_{k-1} - \left( 1/c_k \right) \nabla_\mathbf{x} H(\hat{\mathbf{x}}_{k-1},\hat{\mathbf{y}}_{k-1}),
\end{equation}
and Eq.~\eqref{eq:prox_mag} is equivalent to the ``proximal operator'' 
\begin{equation} \label{eq:PALM_yUpdate}
\begin{split}
\hat{\mathbf{y}}_{k} &= \arg\hspace{-0.5em}\min_{\mathbf{y} \in \mathbb{R}^{N_2}} G(\mathbf{y}) + \frac{d_k}{2} \norm{ \mathbf{y} - \mathbf{z}_k }_2^2,\\
&\triangleq \prox{G}{\mathbf{z}_k,d_k}
\end{split}
\end{equation}
with
\begin{equation}
\mathbf{z}_k = \hat{\mathbf{y}}_{k-1} - \left( 1/d_k \right) \nabla_\mathbf{y} H(\hat{\mathbf{x}}_{k},\hat{\mathbf{y}}_{k-1}).
\end{equation}
These representations are useful, because for many common regularization penalties, the corresponding proximal operators often have simple closed-form solutions \cite{parikh2013, beck2017}.

In summary, the PALM algorithm proceeds according to Alg.~\ref{alg:PALM}.
\begin{algorithm}
	\caption{PALM for Eq.~\eqref{eq:AMcost}}
	\label{alg:PALM}
	\begin{algorithmic}
		\REQUIRE Set $k=1$ and initialize $\hat{\mathbf{x}}_0$ and $\hat{\mathbf{y}}_0$.
		\WHILE {not converge}
		\vspace{0.05in}
		\STATE Choose $c_k \geq L_1(\hat{\mathbf{y}}_{k-1})$\\
		$\mathbf{w}_k = \hat{\mathbf{x}}_{k-1} - \left( 1/c_k \right) \nabla_\mathbf{x} H(\hat{\mathbf{x}}_{k-1},\hat{\mathbf{y}}_{k-1})$\\
		$\hat{\mathbf{x}}_{k} = \prox{F}{\mathbf{w}_k,c_k}$ \\[5pt]
		\STATE Choose $d_k \geq L_2(\hat{\mathbf{x}}_{k})$ \\
		$\mathbf{z}_k = \hat{\mathbf{y}}_{k-1} - \left( 1/d_k \right) \nabla_\mathbf{y} H(\hat{\mathbf{x}}_{k},\hat{\mathbf{y}}_{k-1})$ \\
		$\hat{\mathbf{y}}_{k} = \prox{G}{\mathbf{z}_k,d_k}$ \\[5pt]
		\STATE $k \gets k + 1$
		\ENDWHILE
	\end{algorithmic}
\end{algorithm}

Applying PALM to Eq.~\eqref{eq:GeneralReconModel} is nontrivial and, to the best of our knowledge, has never been done before. We will describe our implementation (as well as the enhancements needed for PALMNUT) in the next section.

\section{Methods} 
This section is organized as follows.  In Subsection~\ref{sec:palm}, we first demonstrate how to apply PALM to solve Eq.~\eqref{eq:GeneralReconModel}. Next, Subsection~\ref{sec:unc} will introduce the principles of our novel approach that relies on uncoupled coordinatewise step sizes. Afterwards, Subsection~\ref{sec:palmnut} will describe the incorporation of Nesterov's momentum strategy and describe the full PALMNUT algorithm. 

For the sake of concreteness, the remainder of this paper will make the further assumption that the phase regularization term $\tilde{R}_2(\mathbf{p})$ can be written as
\begin{equation}\label{eq:PhsRegAssum}
 \tilde{R}_2(\mathbf{p}) = R_2(e^{i\mathbf{p}})
\end{equation}
for some appropriate penalty function $R_2(\cdot)$, such that we are regularizing the exponentiated phase.  This choice is beneficial because it means that $\mathbf{p}$ always appears in exponentiated form in every place it appears in the objective function, which will enable simplifications later on.  This choice has also been used in previous work \cite{JustinDecovSLIM2011,FesslerSepMagPhs2012,haldar2019} because it offers  several other benefits.  First, the exponentiated phase $e^{i\mathbf{p}}$ is unique, even though the phase vector $\mathbf{p}$ is never unique due to the 2$\pi$-periodicity of phase. Second, the exponentiated phase $e^{i\mathbf{p}}$ can be spatially-smooth even if the phase $\mathbf{p}$ is nonsmooth due to issues associated with phase wrapping.  And finally, the spatial derivatives of the exponentiated phase image  can be shown to have the same magnitude as the spatial derivatives of the optimally-unwrapped non-exponentiated phase image under common regularity conditions \cite{ZhiPeiPhaseUnwrap1996}, which is important because phase derivatives are often used by regularization strategies that are designed to promote smooth phase \cite{fessler2004, JustinDecovSLIM2011,FesslerSepMagPhs2012,haldar2019,LustigPhaseCycling2018}.
 
\subsection{Applying PALM to Eq.~\eqref{eq:GeneralReconModel}}\label{sec:palm}
In order to apply PALM to Eq.~\eqref{eq:GeneralReconModel}, it may be tempting to associate $H(\cdot,\cdot)$ with $\tilde{J}(\cdot,\cdot)$, $F(\cdot)$ with $R_1(\cdot)$, and $G(\cdot)$ with $R_2(\cdot)$ as we had also done for AM.  However, it turns out that $\tilde{J}(\cdot,\cdot)$ does not have favorable Lipschitz bounds for the phase subproblem, and a different approach may be preferred.  

To see this, note for the magnitude that 
\begin{equation}
\begin{split}
 &\nabla_{\mathbf{m}} \tilde{J}(\mathbf{m},\mathbf{p}) = \\
 &\hspace{5em}\Re \left\{ e^{-i\mathbf{p}} \odot \left[ \mathbf{A}^{H} \mathbf{A} \left( \mathbf{m} \odot e^{i\mathbf{p}} \right) - \mathbf{A}^{H} \mathbf{b} \right] \right\}
 \end{split}
\end{equation}
and that
\begin{equation}\label{eq:mlip}
\begin{split}
 &\|\nabla_{\mathbf{m}} \tilde{J}(\mathbf{m}_1,\mathbf{p}) - \nabla_{\mathbf{m}} \tilde{J}(\mathbf{m}_2,\mathbf{p})\|_2 \leq \\
 &\hspace{2em}\left\|\Re \left\{ \mathrm{diag}(e^{-i\mathbf{p}})\mathbf{A}^H\mathbf{A} \mathrm{diag}(e^{i\mathbf{p}})\right\} \right\| \|\mathbf{m}_1 - \mathbf{m}_2\|_2,
 \end{split}
\end{equation}
where $\Re\{\cdot\}$ denotes the operator that extracts the real part of its input, $\| \cdot \|$ denotes the spectral norm, and $\mathrm{diag}(e^{i\mathbf{p}})$ is the square diagonal matrix whose diagonal entries are equal to the elements of $ e^{i\mathbf{p}}$.  Due to the characteristics of spectral norms, Eq.~\eqref{eq:mlip} provides a tight Lipschitz bound.

However, for the phase, note that 
\begin{equation}\label{eq:gradp}
 \begin{split}
  &\nabla_{\mathbf{p}} \tilde{J}\left( \mathbf{m,p} \right) = \\
	&\hspace{2em}\Im \left\{ e^{-i\mathbf{p}} \odot \mathbf{m}_{k}\odot \left( \mathbf{A}^{H} \mathbf{A} \left( \mathbf{m}_{k} \odot e^{i\mathbf{p}} \right) - \mathbf{A}^{H} \mathbf{b} \right) \right\},
 \end{split}
\end{equation}
where $\Im\{\cdot\}$ denotes the operator that extracts the imaginary part of its input.  Because of the form of this gradient expression (i.e., $\mathbf{p}$ appears nonlinearly in Eq.~\eqref{eq:gradp}), deriving a good Lipschitz bound for the phase subproblem is nontrivial.  We have been partially successful in deriving valid Lipschitz upper bounds for this case (results not shown), but none of the bounds we've derived have been anywhere close to tight.  This is problematic for the implementation of PALM, because these loose Lipschitz bounds could cause us to choose $d_k$ values that are much larger than necessary, which will result in smaller-than-necessary step sizes in the phase update problem, which will ultimately lead to slow convergence speed.

To avoid this issue, instead of working directly with the original variable $\mathbf{p}$ and its exponentiated version  $e^{i\mathbf{p}}$, we will instead consider an equivalent formulation using the change of variables $\mathbf{q} \triangleq e^{i\mathbf{p}}$ under the constraint that all entries of $\mathbf{q}$ have magnitude one.  This allows us to equivalently express Eq.~\eqref{eq:GeneralReconModel} as
\begin{equation} \label{eq:GeneralReconModel2}
\{\hat{\mathbf{m}},\hat{\mathbf{q}}\} = \arg\min_{\substack{\mathbf{m} \in \mathbb{R}^N \\ \mathbf{q} \in \mathcal{V}}} {J}(\mathbf{m,q}) + R_1(\mathbf{m}) + R_2(\mathbf{q}),
\end{equation}
where 
\begin{equation}\label{eq:F2}
{J}(\mathbf{m,q}) \triangleq \frac{1}{2}\|\mathbf{A}\left(\mathbf{m}\odot \mathbf{q} \right) - \mathbf{b}\|_2^2
\end{equation}
and
\begin{equation}
 \mathcal{V} \triangleq \left\{ \mathbf{q} \in \mathbb{C}^{N}: \abs{q_{n}} = 1,\ n=1,2,...,N \right\}. 
\end{equation}
Once $\hat{\mathbf{m}}$ and $\hat{\mathbf{q}}$ are obtained, the corresponding value of $\hat{\mathbf{p}}$ can be obtained, if so desired, by computing the angle of each of the entries of $\mathbf{q}$.

This reformulation is beneficial, because the gradients simplify substantially.  For the magnitude, we now have that
\begin{equation} \label{eq:DataFid_Mag_Grad}
 \nabla_\mathbf{m} J(\mathbf{m},\mathbf{q}) = \Re\left\{ \bar{\mathbf{q}} \odot \left[\mathbf{A}^H\mathbf{A} (\mathbf{m} \odot \mathbf{q}) - \mathbf{A}^H\mathbf{b} \right] \right\}
\end{equation}
where $\bar{\mathbf{q}}$ denotes the elementwise complex conjugation of $\mathbf{q}$, which gives the tight Lipschitz bound
\begin{equation} \label{eq:DataFid_Mag_Lipschitz}
\begin{split}
 &\|\nabla_{\mathbf{m}} {J}(\mathbf{m}_1,\mathbf{q}) - \nabla_{\mathbf{m}} {J}(\mathbf{m}_2,\mathbf{q})\|_2 \leq \\
 &\hspace{4em}\norm{\Re \left\{ \mathrm{diag}(\bar{\mathbf{q}})\mathbf{A}^H\mathbf{A} \mathrm{diag}(\mathbf{q})\right\}} \|\mathbf{m}_1 - \mathbf{m}_2\|_2.
 \end{split}
\end{equation}
For the phase, we now have that 
\begin{equation} \label{eq:DataFid_Phs_Grad}
 \nabla_\mathbf{q} J(\mathbf{m},\mathbf{q}) =  \mathbf{m} \odot \left[\mathbf{A}^H\mathbf{A} (\mathbf{m} \odot \mathbf{q}) - \mathbf{A}^H\mathbf{b} \right] 
\end{equation}
which gives the tight Lipschitz bound
\begin{equation} \label{eq:DataFid_Phs_Lipschitz}
\begin{split}
 &\|\nabla_{\mathbf{q}} {J}(\mathbf{m},\mathbf{q}_1) - \nabla_{\mathbf{q}} {J}(\mathbf{m},\mathbf{q}_2)\|_2 \leq \\
 &\hspace{8em}\left\| \mathbf{A} \mathrm{diag}(\mathbf{m}) \right\|^2 \|\mathbf{q}_1 - \mathbf{q}_2\|_2.
 \end{split}
\end{equation}

Although this reformulation simplifies the calculation of Lipschitz constants, the introduction of the constraint set $\mathcal{V}$ makes the optimization problem more complicated.  To alleviate this concern, we will further rewrite the constrained problem from Eq.~\eqref{eq:GeneralReconModel2} in an equivalent unconstrained form using indicator functions \cite{AfonsoADMMsparseRecon2010} as
\begin{equation}\label{eq:GeneralReconModel_uncons}
\begin{split}
 \{\hat{\mathbf{m}},\hat{\mathbf{q}}\} = \arg\min_{\substack{\mathbf{m} \in \mathbb{R}^N \\ \mathbf{q} \in \mathbb{C}^N}} {J}(\mathbf{m,q}) &+ R_1(\mathbf{m}) \\
 &+ R_2(\mathbf{q}) + \mathcal{I}_{\mathcal{V}}(\mathbf{q}),
 \end{split}
\end{equation}
where
\begin{equation} 
\mathcal{I_{V}}(\mathbf{q}) \triangleq \begin{cases} 0, & \mathbf{q} \in \mathcal{V} \\ +\infty, & \mathbf{q} \notin \mathcal{V}.	\end{cases}
\end{equation}

The new objective function from Eq.~\eqref{eq:GeneralReconModel_uncons} now has four terms in it  (i.e., $J(\cdot,\cdot)$, $R_1(\cdot)$, $R_2(\cdot)$, and $\mathcal{I}_\mathcal{V}(\cdot)$), and to apply PALM, it is necessary to associate these with the PALM terms $H(\cdot,\cdot)$, $F(\cdot)$, and $G(\cdot)$.  The function $J(\cdot,\cdot)$ is always smooth and involves both $\mathbf{m}$ and $\mathbf{q}$, so we will associate it with $H(\cdot,\cdot)$.  The function $\mathcal{I}_\mathcal{V}(\cdot)$ is always nonsmooth and only involves $\mathbf{q}$, so we will associate it with $G(\cdot)$.  While these associations are straightforward (we have no other options), we potentially have options for $R_1(\cdot)$ and $R_2(\cdot)$ depending on their smoothness characteristics.  If $R_1(\cdot)$ is nonsmooth, then it has to be associated with $F(\cdot)$, but if it is smooth (Lipschitz) then we could either choose to associate it with  $F(\cdot)$ or incorporate it into $H(\cdot,\cdot)$.    Similarly, if $R_2(\cdot)$ is nonsmooth, then it has to be associated with $G(\cdot)$, but if it is smooth (Lipschitz) then we could either choose to associate it with $G(\cdot)$ or incorporate it into $H(\cdot,\cdot)$.  

Although we have different options, the remainder of this paper will assume (for simplicity and without loss of generality) that smooth regularization functions will always be incorporated into $H(\cdot,\cdot)$.  This choice leads to  function associations that are summarized in Table~\ref{tab:summ}.

\bgroup
\def\arraystretch{1.5}
\begin{table}[h]
\resizebox{\columnwidth}{!}{
\begin{tabular}{>{\centering}m{3.5em} >{\centering}m{3.5em} | >{\centering}m{7.5em} | >{\centering}m{3em} | >{\centering\arraybackslash}m{6.5em} |}
$R_1(\cdot)$ is Smooth? & $R_2(\cdot)$ is Smooth? & $H(\cdot,\cdot)$ &  $F(\cdot)$ & $G(\cdot)$ \\ \hline
Y &  Y & $J(\mathbf{m},\mathbf{q}) + R_1(\mathbf{m}) + R_2(\mathbf{q}) $ & 0 & $\mathcal{I}_\mathcal{V}(\mathbf{q})$ \\ \hline
N &  Y & $J(\mathbf{m},\mathbf{q}) + R_2(\mathbf{q})$ & $R_1(\mathbf{m})$ &  $\mathcal{I}_\mathcal{V}(\mathbf{q})$ \\ \hline
Y &  N & $J(\mathbf{m},\mathbf{q}) + R_1(\mathbf{q})$ & 0 & $R_2(\mathbf{q}) + \mathcal{I}_\mathcal{V}(\mathbf{q})$\\ \hline
N &  N & $J(\mathbf{m},\mathbf{q})$ & $R_1(\mathbf{m})$ & $R_2(\mathbf{q}) + \mathcal{I}_\mathcal{V}(\mathbf{q})$ \\ \hline
\end{tabular}}
\caption{Associations between PALM and Eq.~\eqref{eq:GeneralReconModel_uncons} depending on whether $R_1(\cdot)$ and $R_2(\cdot)$ are smooth  (Lipschitz).}
 \label{tab:summ}
\end{table}
\egroup

With these assignments, the PALM algorithm from Alg.~\ref{alg:PALM} can be directly applied, although it is still necessary to specify the computation of $c_k$, $d_k$, $\mathrm{prox}_F(\cdot,\cdot)$, and $\mathrm{prox}_G(\cdot,\cdot)$.  Although these computations will necessarily depend on the characteristics of $R_1(\cdot)$ and $R_2(\cdot)$, we will provide concrete illustrations of these calculations for two typical choices of regularization penalties (we will use these same choices of regularization penalties in the validation study presented later in the paper). Of course, these two illustrations do not encompass every possibility, and interested readers are referred to Refs.~\cite{CombettesProx2011,parikh2013,beck2017} for further discussion and examples of computating proximal operators. However, it should be noted that the two illustrations below focus on the case where $R_2(\cdot)$ is smooth (the first two rows of Table~\ref{tab:summ}), as we have found that it is frequently nontrivial to derive the $\mathrm{prox}_G(\cdot,\cdot)$ operator when $G(\cdot)$ incorporates both $R_2(\cdot)$ and $\mathcal{I}_\mathcal{V}(\cdot)$ (the last two rows of Table~\ref{tab:summ}).

\subsubsection{Huber-function regularization of $\mathbf{m}$ with Tikhonov regularization of $\mathbf{q}$}

For our first illustration, we will consider the case where magnitude regularization takes the form of either
\begin{equation} \label{eq:Huber_Mag_Tikhonov_Phs}
	R_1(\mathbf{m}) = \lambda_1 \sum_{\ell=1}^L h_{\xi}\left(\left[ \mathbf{B} \mathbf{m}\right]_\ell \right)
\end{equation}
or 
\begin{equation} \label{eq:Huber_Mag_Tikhonov_Phs2}
	R_1(\mathbf{m}) = \lambda_1 \sum_{\ell=1}^L h_{\xi}\left( \sqrt{\sum_{t=1}^T \left|\left[\mathbf{B}_{t} \mathbf{m}\right]_\ell\right|^2} \right),
\end{equation}
and phase regularization takes the form of
\begin{equation}\label{eq:r222}
    R_2(\mathbf{q}) = \frac{\lambda_2}{2} \norm{\mathbf{Cq}} _2^2.
\end{equation}
In these expressions, $\lambda_1$ and $\lambda_2$ are positive scalar regularization parameters that can be respectively adjusted to control the influence of the magnitude and phase regularization terms, $[\mathbf{z}]_\ell$ is used to denote the $\ell$th entry of the vector $\mathbf{z}$, and $h_\xi(\cdot)$ is the Huber function defined as 
\begin{equation} \label{eq:Huber}
h_{\xi}(t) = \begin{cases}
\frac{1}{2\xi} \abs{t}^2 , & \abs{t} \leq \xi \\
\abs{t} - \frac{1}{2}\xi, & \abs{t} > \xi.
\end{cases}
\end{equation}

The Huber function is a smooth, convex function that is commonly used for both robust statistics (to mitigate the effects of outliers) \cite{huber1981} and for edge/discontinuity/sparsity-preserving image regularization \cite{nikolova2005,black1995}.  As can be seen from Eq.~\eqref{eq:Huber}, the Huber function is similar to the $\ell_1$-norm for large values of its argument, but unlike the $\ell_1$-norm, is also smooth at the origin because it behaves like a squared $\ell_2$-norm for small values of its argument.  The Huber function with a small value of $\xi$ is frequently chosen as a differentiable surrogate for sparsity-promoting $\ell_1$-norm regularization, while choosing larger values of $\xi$ can make the Huber function more tolerant to smoothly-varying image regions, more resilient to noise, and easier to characterize theoretically \cite{nikolova2005,haldar2013,haldar2011}.  

The regularization in Eq.~\eqref{eq:Huber_Mag_Tikhonov_Phs} is based on applying the Huber function to a single transformation $\mathbf{B}\in\mathbb{C}^{L \times N}$ (e.g., a wavelet transform, a finite-difference transform, etc.) of the magnitude vector $\mathbf{m}$. The more general regularization  in Eq.~\eqref{eq:Huber_Mag_Tikhonov_Phs2} applies the Huber function to a combination of $T$ different transforms $\mathbf{B}_{t}\in\mathbb{C}^{L \times N}$ of $\mathbf{m}$, which can be useful for imposing additional transform-domain structure.    For example, combining a horizontal finite-difference transform with a vertical finite-difference transform within Eq.~\eqref{eq:Huber_Mag_Tikhonov_Phs2} is a common way to achieve isotropic regularization \cite{rudin1991}.  In addition, Eq.~\eqref{eq:Huber_Mag_Tikhonov_Phs2} is related to concepts of simultaneous sparsity \cite{tropp2006}, and our previous work has used regularization penalties of this form to impose the constraint that multi-contrast images of the same anatomy will frequently have correlated edge characteristics \cite{haldar2008,haldar2011,haldar2013,haldar2019}.   Since Eq.~\eqref{eq:Huber_Mag_Tikhonov_Phs} is a special case of Eq.~\eqref{eq:Huber_Mag_Tikhonov_Phs2} with $T=1$, our description below will assume use of the more general form of Eq.~\eqref{eq:Huber_Mag_Tikhonov_Phs2}.

The regularization in Eq.~\eqref{eq:r222} corresponds to standard quadratic (Tikhonov) regularization.  If $\mathbf{C}$ is chosen as a spatial finite-difference operator, this type of regularization can be good at imposing the constraint that the image phase should be spatially smooth without major discontinuities \cite{fessler2004,JustinDecovSLIM2011,haldar2019,FesslerSepMagPhs2012,ZibettiSepMagPhs2017}.  While not every MRI image will have smooth phase characteristics, most do, and smooth phase is a common constraint within the image reconstruction literature \cite{liang1992,haldar2020}.

For this case, both $R_1(\cdot)$ and $R_2(\cdot)$ are smooth, corresponding to the situation in the first row of Table \ref{tab:summ}. As such, to implement PALM, we use the assignments:
\begin{equation}
\begin{split}
 H(\cdot,\cdot) &\leftarrow  \frac{1}{2}\|\mathbf{A}\left(\mathbf{m}\odot \mathbf{q} \right) - \mathbf{b}\|_2^2 \\ 
 & \hspace{0.2in}+  \lambda_1 \sum_{\ell=1}^L h_{\xi}\left( \sqrt{\sum_{t=1}^T \left|\left[\mathbf{B}_{t} \mathbf{m}\right]_\ell\right|^2} \right) \\
 &\hspace{0.2in}+  \lambda_2 \norm{\mathbf{Cq}} _2^2 \\
 F(\cdot) &\leftarrow 0 \\
 G(\cdot) &\leftarrow \mathcal{I}_\mathcal{V}(\mathbf{q}).
 \end{split}
\end{equation}

The gradients of $H(\cdot,\cdot)$ needed for PALM are
\begin{equation}
 \begin{split}
  \nabla_{\mathbf{m}}H(\mathbf{m},\mathbf{q}) = \Re&\left\{ \bar{\mathbf{q}} \odot \left[\mathbf{A}^H\mathbf{A} (\mathbf{m} \odot \mathbf{q}) - \mathbf{A}^H\mathbf{b} \right] \right. \\
  &+ \left. \lambda_1 \sum_{t=1}^T \mathbf{B}_t^H \mathbf{W}(\mathbf{m}) \mathbf{B}_t \mathbf{m} \right\}
 \end{split}
\end{equation}
and
\begin{equation}
\begin{split}
 \nabla_{\mathbf{q}}H(\mathbf{m},\mathbf{q}) =\mathbf{m} \odot& \left[\mathbf{A}^H\mathbf{A} (\mathbf{m} \odot \mathbf{q}) - \mathbf{A}^H\mathbf{b} \right] \\
 &+\lambda_2 \mathbf{C}^H\mathbf{C}\mathbf{q},
 \end{split}
\end{equation}
where $\mathbf{W}(\mathbf{m})$ is an $L\times L$ diagonal matrix depending on $\mathbf{m}$ with $\ell$th diagonal entry given by
\begin{equation}
 [\mathbf{W}(\mathbf{m})]_{\ell \ell} =  1/\max\left\{\xi,\sqrt{\sum_{t=1}^T \left|\left[\mathbf{B}_{t} \mathbf{m}\right]_\ell\right|^2} \right\}.
\end{equation}

These gradient expressions give rise to Lipschitz-type upper bounds in the form of Eqs.~\eqref{eq:PALM_Lipx} and \eqref{eq:PALM_Lipy}, such that the majorization and descent conditions will be satisfied whenever
\begin{equation}
 c_k \geq \norm{\mathbf{A}}^2+\frac{\lambda_1}{\xi} \norm{\sum_{t=1}^T \mathbf{B}_t^H\mathbf{B}_t}
\end{equation}
and
\begin{equation}\label{eq:d1}
 d_k \geq \norm{\mathbf{A}}^2 \norm{\hat{\mathbf{m}}_{k}}_\infty^2+\lambda_2 \norm{\mathbf{C}}^2,
\end{equation}
where $\|\cdot\|_\infty$ denotes the $\ell_\infty$-norm. The spectral norms in these expressions are not iteration-dependent, and can be precomputed and reused throughout the iterative process.  If it is difficult to analytically calculate the spectral norm values, they can also be evaluated using standard computationally-efficient numerical methods like Lanczos iteration \cite{golub2013}.

Finally, the proximal operators needed for PALM are given by
\begin{equation}
\prox{F}{\mathbf{w}_k,c_k} = \mathbf{w}_k
\end{equation}
and
\begin{equation}\label{eq:proxg}
\prox{G}{\mathbf{z}_k,d_k} = \frac{\mathbf{z}_k}{|\mathbf{z}_k|},
\end{equation}
where division is performed elementwise and we choose the convention that $\frac{0}{0}=1$.

\subsubsection{$\ell_1$ regularization of $\mathbf{m}$ with Huber-function regularization of $\mathbf{q}$}  

For our second illustration, we will consider the case where 
\begin{equation} \label{eq:L1_Mag_Huber_Phs}
	R_1(\mathbf{m}) = \lambda_1 \norm{\mathbf{Tm}}_1
	\end{equation}
	and
	\begin{equation}\label{eq:L1_Mag_Huber_Phs2}
	R_2(\mathbf{q}) = \lambda_2 \sum_{\ell=1}^L h_{\xi}\left( \sqrt{\sum_{t=1}^T \left|\left[\mathbf{B}_{t} \mathbf{q}\right]_\ell\right|^2} \right).
\end{equation}
The $\ell_1$-norm penalty with sparsifying transform matrix $\mathbf{T}$ from Eq.~\eqref{eq:L1_Mag_Huber_Phs} is standard for promoting transform-domain sparsity, and has been previously used to regularize the magnitude vector $\mathbf{m}$ in several applications \cite{cetin2001,tuysuzoglu2012,FesslerSepMagPhs2012,guven2016,ZibettiSepMagPhs2017,LustigPhaseCycling2018,moradikia2018}. The characteristics of the Huber function from Eq.~\eqref{eq:L1_Mag_Huber_Phs2} have been discussed previously.  By taking a small value of the parameter $\xi$, the Huber function can be used as a smooth approximation of the $\ell_1$-norm  in order to enable sparsity-promoting and/or edge-preserving regularization of the phase image, which can useful for some applications with more complicated phase characteristics \cite{LustigPhaseCycling2018, FesslerSepMagPhs2012}.

In this case, $R_1(\cdot)$ is non-smooth while $R_2(\cdot)$ is smooth, corresponding to the second row of Table \ref{tab:summ}.  As such, to implement PALM, we use the assignments:
\begin{equation}
\begin{split}
 H(\cdot,\cdot) &\leftarrow  \frac{1}{2}\|\mathbf{A}\left(\mathbf{m}\odot \mathbf{q} \right) - \mathbf{b}\|_2^2 \\ 
 & \hspace{0.2in}+  \lambda_2 \sum_{\ell=1}^L h_{\xi}\left( \sqrt{\sum_{t=1}^T \left|\left[\mathbf{B}_{t} \mathbf{q}\right]_\ell\right|^2} \right) \\
 F(\cdot) &\leftarrow \lambda_1 \norm{\mathbf{Tm}}_1 \\
 G(\cdot) &\leftarrow \mathcal{I}_\mathcal{V}(\mathbf{q}).
 \end{split}
\end{equation}

The gradients of $H(\cdot,\cdot)$ needed for PALM are
\begin{equation}
 \begin{split}
  \nabla_{\mathbf{m}}H(\mathbf{m},\mathbf{q}) = \Re&\left\{ \bar{\mathbf{q}} \odot \left[\mathbf{A}^H\mathbf{A} (\mathbf{m} \odot \mathbf{q}) - \mathbf{A}^H\mathbf{b} \right] \right\}
 \end{split}
\end{equation}
and
\begin{equation}
\begin{split}
 \nabla_{\mathbf{q}}H(\mathbf{m},\mathbf{q}) =\mathbf{m} \odot& \left[\mathbf{A}^H\mathbf{A} (\mathbf{m} \odot \mathbf{q}) - \mathbf{A}^H\mathbf{b} \right] \\
  &+ \lambda_2 \sum_{t=1}^T \mathbf{B}_t^H \mathbf{W}(\mathbf{q}) \mathbf{B}_t \mathbf{q}.
 \end{split}
\end{equation}

These gradient expressions give rise to Lipschitz-type upper bounds in the form of Eqs.~\eqref{eq:PALM_Lipx} and \eqref{eq:PALM_Lipy}, such that the majorization and descent conditions will be satisfied whenever
\begin{equation}
 c_k \geq \norm{\mathbf{A}}^2
\end{equation}
and
\begin{equation}\label{eq:d2}
 d_k \geq \norm{\mathbf{A}}^2 \norm{\hat{\mathbf{m}}_{k}}_\infty^2 + \frac{\lambda_2}{\xi} \norm{\sum_{t=1}^T \mathbf{B}_t^H\mathbf{B}_t}.
\end{equation}
As before, the spectral norms in these expressions are not iteration-dependent, and can be precomputed and reused throughout the iterative process.

Finally, assuming that $\mathbf{T}$ is a unitary transform such that $\mathbf{T}^H = \mathbf{T}^{-1}$, the proximal operator for $F(\cdot)$ is given by \cite{beck2017}
\begin{equation} \label{eq:SoftThr}
\begin{split}
\prox{F}{\mathbf{w}_k,c_k} &= \\
&\hspace{-2em}\mathbf{T}^H\mathrm{diag}\left(\frac{\max\left\{|\mathbf{T}\mathbf{w}_k| - \lambda_1/c_k,0\right\}}{|\mathbf{T}\mathbf{w}_k|}\right)\mathbf{T}\mathbf{w}_k,
\end{split}
\end{equation}
where maximization, absolute value, and division operations are performed elementwise.  Note that $G(\cdot)$ is the same as in the previous illustration, and therefore has the same proximal operator (Eq.~\eqref{eq:proxg}).

\subsection{Uncoupled Step Sizes}\label{sec:unc}

Although the PALM algorithm described in the previous section provides a novel and effective approach for solving Eq.~\eqref{eq:GeneralReconModel}, we observe in this section that it may be very conservative and computationally inefficient for PALM to use the same value of $d_k$ (and therefore, the same step size $1/d_k$) for all elements of the phase vector $\mathbf{p}$.  This inefficiency stems from the fact that $d_k$ is set based on the global Lipschitz constant (effectively, the worst-case rate of change of the gradient along any possible direction), while we have observed that the rate of change of the gradient can be much smaller than the worst-case along specific directions.  Concretely, using the global Lipschitz constant means that the step size will depend on the maximum value of $\mathbf{m}$, while we observe that it can be much better for the step size for each coordinate to instead depend on the coordinatewise  values of $\mathbf{m}$.  This observation motivates us to investigate and utilize coordinatewise bounds on the rate of change of the gradient, enabling uncoupled coordinatewise step sizes.  

For the sake of generality, we will first describe this approach for the general setting of Section~\ref{sec:back}, where we are given a generic smooth real-valued function $H(\mathbf{x},\mathbf{y})$. The PALM approach utilized majorants of $H(\mathbf{x},\mathbf{y})$ that were derived based on the global (scalar-valued) Lipschitz constants $L_1(\mathbf{y})$ and $L_2(\mathbf{x})$ of Eqs.~\eqref{eq:PALM_Lipx} and \eqref{eq:PALM_Lipy}.  In this section, we instead make the assumption that a vector $\mathbf{L}_1(\mathbf{y}) \in \mathbb{R}^{N_1}$  can be found such that, for a fixed value of $\mathbf{y}$, we have 
\begin{equation} \label{eq:Lipd1}
\begin{split}
\inner{ \nabla_\mathbf{x} H(\mathbf{x}_1,{\mathbf{y}}) - \nabla_\mathbf{x} H(\mathbf{x}_2,{\mathbf{y}}), \mathbf{x}_1-\mathbf{x}_2 } & \\
& \hspace{-6em} \leq  \norm{\sqrt{\mathbf{L}_1(\mathbf{y})} \odot (\mathbf{x}_1 - \mathbf{x}_2) }_2^2
\end{split}
\end{equation}
for $\forall \mathbf{x}_1,\mathbf{x}_2 \in \mathbb{R}^N_{1}$, where the square-root operation is applied elementwise.  Similarly, we assume that a vector $\mathbf{L}_2(\mathbf{x}) \in \mathbb{R}^{N_2}$ can be found such that, for a fixed value of $\mathbf{x}$, we have
\begin{equation}\label{eq:Lipd2}
\begin{split}
\inner{ \nabla_\mathbf{y} H(\mathbf{x},{\mathbf{y}_1}) - \nabla_\mathbf{y} H(\mathbf{x},{\mathbf{y}_2}), \mathbf{y}_1-\mathbf{y}_2 } & \\
& \hspace{-6em} \leq  \norm{\sqrt{\mathbf{L}_2(\mathbf{x})} \odot (\mathbf{y}_1 - \mathbf{y}_2) }_2^2
\end{split}
\end{equation}	
for $\forall \mathbf{y}_1,\mathbf{y}_2 \in \mathbb{R}^{N_2}$.  

It should be noted that if the global Lipschitz continuity conditions of Eqs.~\eqref{eq:PALM_Lipx} and \eqref{eq:PALM_Lipy} are known to be satisfied, then a vector $\mathbf{L}_1(\mathbf{y})$ satisfying Eq.~\eqref{eq:Lipd1} can be trivially obtained by setting all of its entries equal to $L_1(\mathbf{y})$, with an analogous argument holding true for  $\mathbf{L}_2(\mathbf{x})$.  In particular,  the Cauchy-Schwarz inequality combined with Eq.~\eqref{eq:PALM_Lipx}  implies that 
\begin{equation}
\begin{split}
 \inner{ \nabla_\mathbf{x} H(\mathbf{x}_1,{\mathbf{y}}) - \nabla_\mathbf{x} H(\mathbf{x}_2,{\mathbf{y}}), \mathbf{x}_1-\mathbf{x}_2 } &\\
 & \hspace{-14em} \leq \norm{\nabla_\mathbf{x} H(\mathbf{x}_1,{\mathbf{y}}) - \nabla_\mathbf{x} H(\mathbf{x}_2,{\mathbf{y}})}_2 \norm{\mathbf{x}_1-\mathbf{x}_2}_2\\
 & \hspace{-14em} \leq L_1(\mathbf{y})\norm{\mathbf{x}_1-\mathbf{x}_2}_2^2\\
 & \hspace{-14em}= \norm{\sqrt{L_1(\mathbf{y})}(\mathbf{x}_1-\mathbf{x}_2)}_2^2.
 \end{split}
\end{equation}
However, Eqs.~\eqref{eq:Lipd1} and \eqref{eq:Lipd2} are more flexible than Eqs.~\eqref{eq:PALM_Lipx} and \eqref{eq:PALM_Lipy} because a different Lipschitz-like constant can be used for every coordinate, and many of these entries can be potentially much smaller than the global Lipschitz constant (because the function gradient may change much more slowly along these directions).

From an optimization perspective, Eqs.~\eqref{eq:Lipd1} and \eqref{eq:Lipd2} are important because they enable the use of potentially better majorants than were used by PALM, as described by the following theorem.  
\begin{theorem} \label{thm:UT_Conv}
Consider the setting described in Section~\ref{sec:back}, and assume that the smooth real-valued function $H(\mathbf{x},\mathbf{y})$ satisfies the conditions of Eq.~\eqref{eq:Lipd1}.  Then given a vector  $\mathbf{c}_k \in \mathbb{R}^{N_1}$ and assuming $\mathbf{y}$ is held fixed at $\mathbf{y} = \hat{\mathbf{y}}_{k-1}$, the function
\begin{equation}
\begin{split}
 H(&\hat{\mathbf{x}}_{k-1},\hat{\mathbf{y}}_{k-1}) + \inner{\mathbf{x}-\hat{\mathbf{x}}_{k-1}, \nabla_\mathbf{x} H(\hat{\mathbf{x}}_{k-1},\hat{\mathbf{y}}_{k-1})} \\ &+ \frac{1}{2} \norm{\sqrt{\mathbf{c}_k}\odot(\mathbf{x} - \hat{\mathbf{x}}_{k-1})}_2^2 + F(\mathbf{x}) + G(\hat{\mathbf{y}}_{k-1})
 \end{split}
\end{equation}
is a majorant of $\Psi(\mathbf{x},\hat{\mathbf{y}}_{k-1})$ at the point $\mathbf{x}=\hat{\mathbf{x}}_{k-1}$ whenever  $\mathbf{c}_k \geq \mathbf{L}_1(\hat{\mathbf{y}}_{k-1})$ (elementwise).  Similarly, given a vector $\mathbf{d}_k \in \mathbb{R}^{N_2}$, assuming $H(\mathbf{x},\mathbf{y})$ satisfies the conditions of Eq.~\eqref{eq:Lipd2}, and assuming $\mathbf{x}$ is held fixed at $\mathbf{x} = \hat{\mathbf{x}}_{k}$, the function 
\begin{equation}
\begin{split}
 H(&\hat{\mathbf{x}}_{k},\hat{\mathbf{y}}_{k-1}) + \inner{\mathbf{y}-\hat{\mathbf{y}}_{k-1}, \nabla_\mathbf{y} H(\hat{\mathbf{x}}_{k},\hat{\mathbf{y}}_{k-1})} \\ &+ \frac{1}{2} \norm{\sqrt{\mathbf{d}_k}\odot(\mathbf{y} - \hat{\mathbf{y}}_{k-1})}_2^2 + F(\hat{\mathbf{x}}_{k}) + G(\mathbf{y})
 \end{split}
\end{equation}
is a majorant of $\Psi(\hat{\mathbf{x}}_{k},\mathbf{y})$ at the point $\mathbf{y} = \hat{\mathbf{y}}_{k-1}$ whenever $\mathbf{d}_k \geq \mathbf{L}_2(\hat{\mathbf{x}}_{k})$ (elementwise).
\end{theorem}
The proof of this theorem is given in Appendix~\ref{app:proof}. Although this theorem is stated for real-valued vectors (for consistency with previous descriptions), the same also holds true for complex-valued vectors $\mathbf{x}$ and $\mathbf{y}$.

Following the same approach as PALM but replacing the original PALM majorants (Eqs.~\eqref{eq:prox_phs} and \eqref{eq:prox_mag}) with the new majorants from Thm.~\ref{thm:UT_Conv} results in the new PALM algorithm with uncoupled step sizes given in Alg.~\ref{alg:PALMUT}. This algorithm uses proximal operators with vector-valued $\mathbf{c}_k$ and $\mathbf{d}_k$, which we define as
\begin{equation}
\begin{split}
 \prox{F}{\mathbf{w}_k,\mathbf{c}_k} \triangleq \arg\hspace{-0.5em}\min_{\mathbf{x} \in \mathbb{R}^{N_1}} &F(\mathbf{x}) \\&+ \frac{1}{2} \norm{\sqrt{\mathbf{c}_k} \odot( \mathbf{x} - \mathbf{w}_k )}_2^2
 \end{split}
\end{equation}
and
\begin{equation} 
\begin{split}
\prox{G}{\mathbf{z}_k,\mathbf{d}_k} \triangleq  \arg\hspace{-0.5em}\min_{\mathbf{y} \in \mathbb{R}^{N_2}} &G(\mathbf{y}) \\&+ \frac{1}{2} \norm{ \sqrt{\mathbf{d}_k} \odot (\mathbf{y} - \mathbf{z}_k) }_2^2.
 \end{split}
\end{equation}

\begin{algorithm}
	\caption{PALM for Eq.~\eqref{eq:AMcost} with Uncoupled Step Sizes}
	\label{alg:PALMUT}
	\begin{algorithmic}
		\REQUIRE Set $k=1$ and initialize $\hat{\mathbf{x}}_0$ and $\hat{\mathbf{y}}_0$.
		\WHILE {not converge}
		\vspace{0.05in}
		\STATE Choose $\mathbf{c}_k \geq \mathbf{L}_{1}(\hat{\mathbf{y}}_{k-1})$ (elementwise)\\
		$\mathbf{w}_k = \hat{\mathbf{x}}_{k-1} - \text{diag}(\mathbf{c}_k)^{-1} \nabla_\mathbf{x} H(\hat{\mathbf{x}}_{k-1},\hat{\mathbf{y}}_{k-1})$\\
		$\displaystyle \hat{\mathbf{x}}_{k} = \prox{F}{\mathbf{w}_k,\mathbf{c}_k}$ \\[5pt]
		\STATE Choose $\mathbf{d}_k \geq \mathbf{L}_{2}(\hat{\mathbf{x}}_{k})$ (elementwise) \\
		$\mathbf{z}_k = \hat{\mathbf{y}}_{k-1} - \text{diag}(\mathbf{d}_k)^{-1} \nabla_\mathbf{y} H(\hat{\mathbf{x}}_{k},\hat{\mathbf{y}}_{k-1})$ \\
		$\displaystyle \hat{\mathbf{y}}_{k} = \prox{G}{\mathbf{z}_k,\mathbf{d}_k}$ \\[5pt]
		\STATE $k \gets k + 1$
		\ENDWHILE
	\end{algorithmic}
\end{algorithm}

Just like for the original PALM algorithm, the use of valid majorants in our new algorithm guarantees that it will monotonically decrease the objective function value, and that the objective function value converge if $\Psi(\mathbf{x},\mathbf{y})$ is bounded from below.  Further, this new algorithm reduces to the original PALM algorithm if the $\mathbf{c}_k$ and $\mathbf{d}_k$ are treated as scalars (i.e., the values in different entries are always the same) instead of choosing different values in each of the entries.

Now that the general approach has been described, let's consider the application  to the specific magnitude and phase optimization problem of interest from Eq.~\eqref{eq:GeneralReconModel}.  Without further assumptions about the structure of $\mathbf{A}$, we do not observe any special coordinatewise structure for the magnitude subproblem.  As such, we can simply utilize the global Lipschitz constant in this case, setting $\mathbf{c}_k = c_k \mathbf{1}$, where $\mathbf{1}$ is a vector whose entries are all equal to 1 and $c_k$ is the value obtained for the PALM algorithm as described in the previous section.

However, for the phase subproblem with a fixed value of $\mathbf{m}$, we observe that 
\begin{equation} 
\begin{split}
 &\inner{\nabla_{\mathbf{q}} {J}(\mathbf{m},\mathbf{q}_1) - \nabla_{\mathbf{q}} {J}(\mathbf{m},\mathbf{q}_2), \mathbf{q}_1 - \mathbf{q}_2} \leq \\
 &\hspace{8em}\|\norm{\mathbf{A}} |\mathbf{m}| \odot (\mathbf{q}_1 - \mathbf{q}_2)\|_2^2,
 \end{split}
\end{equation}
where the magnitude operation is applied elementwise to the vector $\mathbf{m}$.  Thus, for the first illustration from the previous subsection (Huber-function regularization of $\mathbf{m}$ with Tikhonov regularization of $\mathbf{p}$), our new algorithm can adopt
\begin{equation}
 \mathbf{d}_k \geq \norm{\mathbf{A}}^2 |\hat{\mathbf{m}}_k|^2 + \lambda_2 \norm{\mathbf{C}}^2
\end{equation}
instead of the previous expression from Eq.~\eqref{eq:d1}.  Similarly, for the second illustration from the previous subsection ($\ell_1$-regularization of $\mathbf{m}$ with Huber-function regularization of $\mathbf{p}$), our new algorithm can adopt
\begin{equation}
 \mathbf{d}_k \geq \norm{\mathbf{A}}^2 |\hat{\mathbf{m}}_k|^2 +\frac{\lambda_2}{\xi} \norm{\sum_{t=1}^T \mathbf{B}_t^H\mathbf{B}_t}
\end{equation}
instead of the previous expression from Eq.~\eqref{eq:d2}.  Note also that for both of the previous illustrations, the $\prox{G}{\cdot,\cdot}$ expressions had no dependence on the actual value of $d_k$, which allows us to simply reuse the same proximal operators for the new algorithm without modification. As such, for these illustrations, the main difference between the PALM algorithm and our new algorithm is that PALM takes a uniform step size for the phase update that depends on the maximium value of $\mathbf{m}$, while the new algorithm can take larger step sizes for coordinates where the corresponding value of $\mathbf{m}$ is small.

\subsection{Nesterov's Momentum Acceleration and PALMNUT}\label{sec:palmnut}
Our proposed  PALMNUT (PALM with Nesterov's momentum and Uncoupled sTep sizes) algorithm is obtained by combining Alg.~\ref{alg:PALMUT} with Nesterov's momentum technique.  The basic idea of Nesterov's technique is that, instead of computing the next iterate $\hat{\mathbf{x}}_{k}$ based on values derived from  $\hat{\mathbf{x}}_{k-1}$, it can be better to instead find the next iterate using values derived from a combination of the previous iterates  \cite{NesterovBook2004,BeckFISTA2009}, which can be interpreted as using ``momentum'' from previous iterations.  For convex optimization problems, this approach can even result in convergence rates that have optimal order \cite{NesterovBook2004,BeckFISTA2009}.  Of course, the problem of interest in this work is not convex, although it has been shown empirically (oftentimes without rigorous theoretical justification) that Nesterov's momentum technique  can often substantially accelerate the iterative solution of nonconvex optimization problems.

The idea of applying momentum to accelerate the convergence of PALM has been studied in Ref.~\cite{PockiPALM2016}, where the resulting algorithm was called inertial PALM (iPALM).  Theoretical convergence results for iPALM were proven with restrictive parameter choices (different from Nesterov's original parameter choices), although it was also shown empirically that using Nesterov's original parameter choices generally led to much faster convergence, despite the lack of theoretical guarantees.  Our empirical experience is also consistent with these previous observations, so our proposed PALMNUT algorithm similarly utilizes Nesterov's original parameter choices (following the concise form described by Ref.~\cite{SuODE2014}).  The final PALMNUT algorithm, incorporating both uncoupled coordinatewise step sizes and Nesterov's momentum, is given in Alg.~\ref{alg:PALMNUT}.

\begin{algorithm}
	\caption{PALMNUT}
	\label{alg:PALMNUT}
	\begin{algorithmic}
		\REQUIRE Set $k=1$ and initialize $\hat{\mathbf{x}}_0$ and $\hat{\mathbf{y}}_0$.  \\
		Set $\mathbf{u}_{0}=\hat{\mathbf{x}}_{0}$ and $\mathbf{v}_{0}=\hat{\mathbf{y}}_{0}$.
		\WHILE {not converge}
		\vspace{0.05in}
		\STATE Choose $\mathbf{c}_k \geq \mathbf{L}_1(\mathbf{v}_{k-1})$ (elementwise) \\
		$\mathbf{w}_k = \hat{\mathbf{x}}_{k-1} - \text{diag}(\mathbf{c}_k)^{-1} \nabla_\mathbf{x} H({\mathbf{u}}_{k-1},\mathbf{v}_{k-1})$\\
		$\displaystyle \hat{\mathbf{x}}_{k} = \prox{F}{\mathbf{w}_k,\mathbf{c}_k}$ \\[5pt]
		$\mathbf{u}_{k} = \hat{\mathbf{x}}_{k} + \frac{k-1}{k+2}(\hat{\mathbf{x}}_{k} -\hat{\mathbf{x}}_{k-1})$\\[5pt]
		\STATE Choose $\mathbf{d}_k \geq \mathbf{L}_{2}({\mathbf{u}}_{k})$ (elementwise) \\
		$\mathbf{z}_k = \hat{\mathbf{y}}_{k-1} - \text{diag}(\mathbf{d}_k)^{-1} \nabla_\mathbf{y} H(\mathbf{u}_{k},{\mathbf{v}}_{k-1})$ \\
		$\displaystyle \hat{\mathbf{y}}_{k} = \prox{G}{\mathbf{z}_k,\mathbf{d}_k}$ \\[5pt]		
		$\mathbf{v}_{k} = \hat{\mathbf{y}}_{k} + \frac{k-1}{k+2}(\hat{\mathbf{y}}_{k} -\hat{\mathbf{y}}_{k-1})$\\[5pt]
		\STATE $k \gets k + 1$
		\ENDWHILE
	\end{algorithmic}
\end{algorithm}

\section{Numerical Experiments}

In the following subsections, we evaluate PALMNUT in three different simulations that are representative of a diverse set of real problems in MRI: sparsity-promoting reconstruction of undersampled k-space data \cite{FesslerSepMagPhs2012,ZibettiSepMagPhs2017,LustigPhaseCycling2018}, regularization-based denoising of a complex image \cite{haldar2019,haldar2013, haldar2008}, and using phase correction to enable the combination of multiple images acquired in the presence of experimental phase instabilities \cite{haldar2019, bernstein1989,mckinnon2000,liu2004,ChenMUSE2013,eichner2015}.

In each of these cases, we compare PALMNUT against AM combined with NCG  \cite{fessler2004,JustinDecovSLIM2011,FesslerSepMagPhs2012, ZibettiSepMagPhs2017,haldar2019} and AM combined with phase cycling \cite{LustigPhaseCycling2018}.  PALMNUT and AM combined with NCG are directly comparable, because they can both be used to regularize the exponentiated phase $e^{i\mathbf{p}}$, and therefore can both be applied to the exact same optimization problem.  As a result, for both of these algorithms, the phase was regularized with $R_2(e^{i\mathbf{p}})$ as described previously, and we used the cost function value and the computation time to judge algorithm performance.  However, the phase cycling heuristic \cite{LustigPhaseCycling2018} is intended to be used for the direct regularization of $\mathbf{p}$.  For this algorithm, we therefore had to instead use  phase regularization of the form $R_2(\mathbf{p})$.  In addition, by its nature, the phase cycling heuristic employs a different phase-cycled cost function at each iteration, which makes it difficult to plot a meaningful cost function value.   To allow for comparisons with this different approach, we therefore also computed a normalized root-mean-squared error (NRMSE) metric for each algorithm.  If $\mathbf{f}$ represents the ground truth vector and $\hat{\mathbf{f}}$ represents an estimate, the NRMSE of the estimate is given by
\begin{equation}
 \mathrm{NRMSE} \triangleq \|\hat{\mathbf{f}}-\mathbf{f}\|_2/\|\mathbf{f}\|_2.
\end{equation}
In order to focus on consequential errors, the NRMSE values were always computed after masking out the empty (background) parts of the field-of-view. This was particularly beneficial for AM with phase cycling, which frequently showed higher error levels in the image background compared to the other two approaches. 

Regularization parameters for each cost function in each scenario were empirically optimized to achieve the smallest possible final NRMSE values for the two AM algorithms.  Our implementation of AM with NCG used the Polak-Ribiere version of NCG \cite{press1992}. Our implementation of AM with phase cycling was based on code provided by the authors of Ref.~\cite{LustigPhaseCycling2018} (available from \url{https://mrirecon.github.io/bart/}).

\begin{figure}[t]
	\centering
	\subfloat[Magnitude]{\includegraphics[height=1.12in]{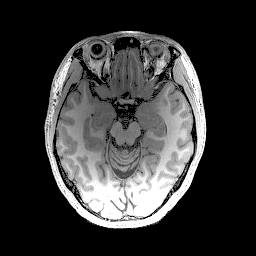}}
	\hfil
	\subfloat[Phase]{\includegraphics[height=1.12in]{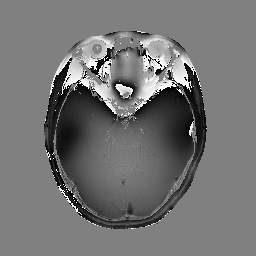}}
	\hfil
\subfloat[Sampling Mask]{\includegraphics[height=1.12in]{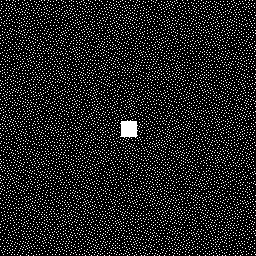}}
	\caption{The ground truth (a) magnitude and (b) phase images used for the undersampled MRI reconstruction scenario, along with (c) the $8\times$-accelerated k-space sampling mask used to simulate the undersampled acquisition.}\label{fig:goldstandard}
\end{figure}

\subsection{Undersampled MRI Reconstruction}\label{sec:under}

In the first set of evaluations, we considered the reconstruction of an MR image from $8\times$-undersampled k-space data.  The gold-standard magnitude and phase images, which were obtained from a real fully-sampled in vivo T1-weighted MRI acquisition with $256 \times 256$ in-plane matrix size, are shown in Fig.~\ref{fig:goldstandard}. This figure also shows the k-space sampling mask (corresponding to $8\times$ undersampling) that we used to simulate an accelerated acquisition.

\begin{figure}
	\centering
	\subfloat[]{\includegraphics[width=1.7in]{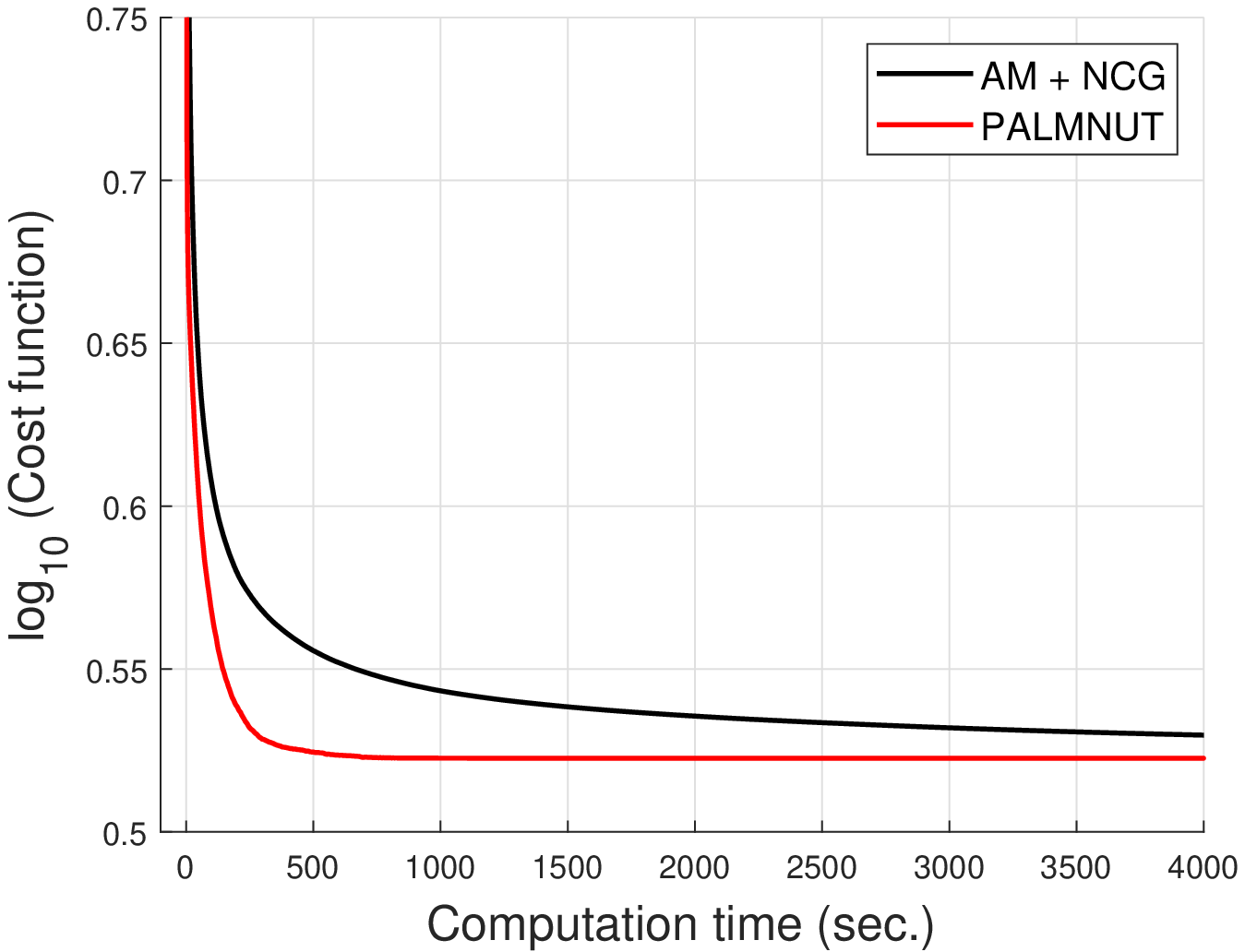}}
	\hfil
	\subfloat[]{\includegraphics[width=1.7in]{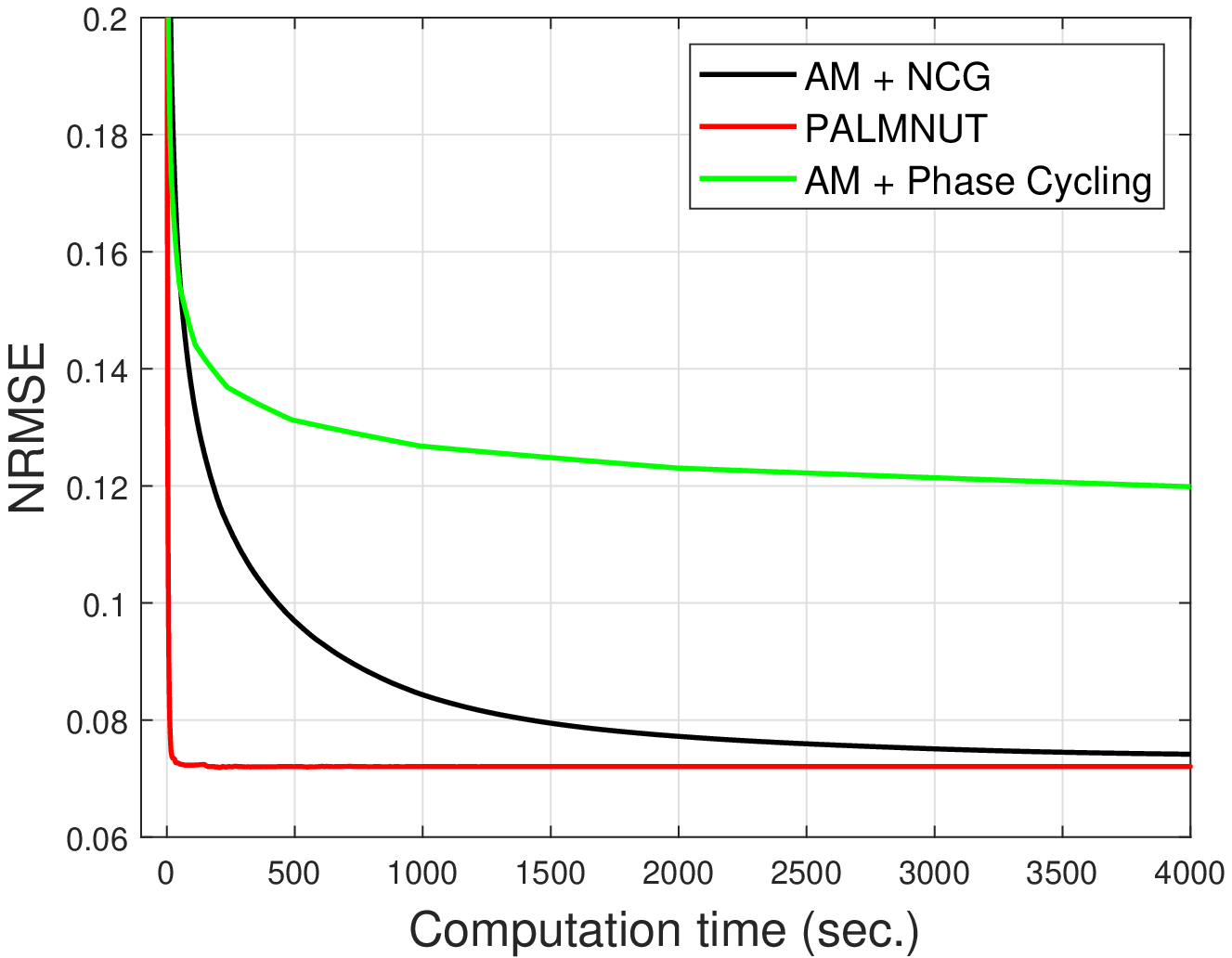}}\\
	\subfloat[Zero-filled]{\includegraphics[width=.8in]{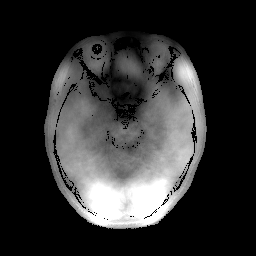}\includegraphics[width=.8in]{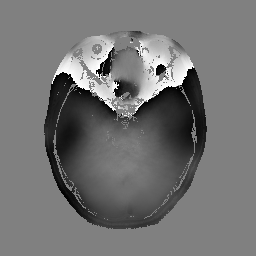}}
	\hfil
	\subfloat[PALMNUT]{\includegraphics[width=.8in]{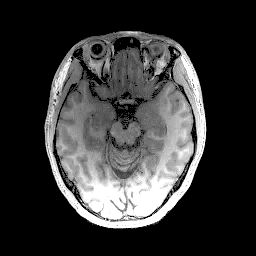}\includegraphics[width=.8in]{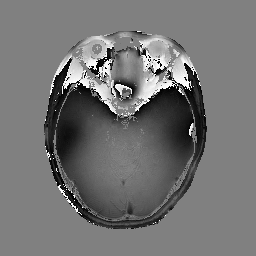}}
	\caption{Convergence plots and reconstructed images for the undersampled MRI reconstruction scenario. The convergence plots show (a) the cost function value as a function of computation time and (b) the NRMSE value as a function of computation time.    Also shown are the magnitude and phase images corresponding to (c) zero-filled reconstruction and (d) PALMNUT.}
	\label{fig:CSpMRIConvCurv}
\end{figure}

For this simulation, the original fully-sampled k-space data (originally measured with 32 channels) was coil-compressed down to 8 virtual channels to reduce computational complexity, and was then retrospectively undersampled using the aforementioned k-space sampling mask.  For reconstruction, the $\mathbf{A}$ matrix was chosen according to the standard SENSE model \cite{pruessmann2001}, with sensitivity maps estimated using ESPIRiT \cite{LustigESPIRiT}. The magnitude regularization took the form of an $\ell_1$ penalty as given by Eq.~\eqref{eq:L1_Mag_Huber_Phs}, where, following Ref.~\cite{LustigPhaseCycling2018}, the sparsifying transform $\mathbf{T}$ was chosen to be the unitary Daubechies-4 wavelet transform.  The phase regularization took the form of a Huber-function penalty as given by Eq.~\eqref{eq:L1_Mag_Huber_Phs2}, where the Huber parameter $\xi$ was chosen to be a small number (i.e., $\xi = 0.001$) in order to approximate the $\ell_1$-norm.  Following Ref.~\cite{LustigPhaseCycling2018}, the transform we used for phase regularization was also a unitary Daubechies-4 wavelet transform.  All three algorithms were initialized by applying SENSE-based coil-combination to the multi-channel images obtained by zero-filling the unmeasured data.

Convergence results for all three algorithms are shown in Fig.~\ref{fig:CSpMRIConvCurv}, along with representative image reconstruction results.  As can be seen, the cost function and NRMSE values converge to similar levels for both  PALMNUT and AM with NCG, although PALMNUT converged much faster.  Although AM with phase cycling converged to a result with a reasonably-good NRMSE, it was substantially worse than PALMNUT in both convergence speed and the final achieved NRMSE value.

\begin{figure}[t]
	\centering
	\subfloat[Ground Truth]{\includegraphics[width=.85in]{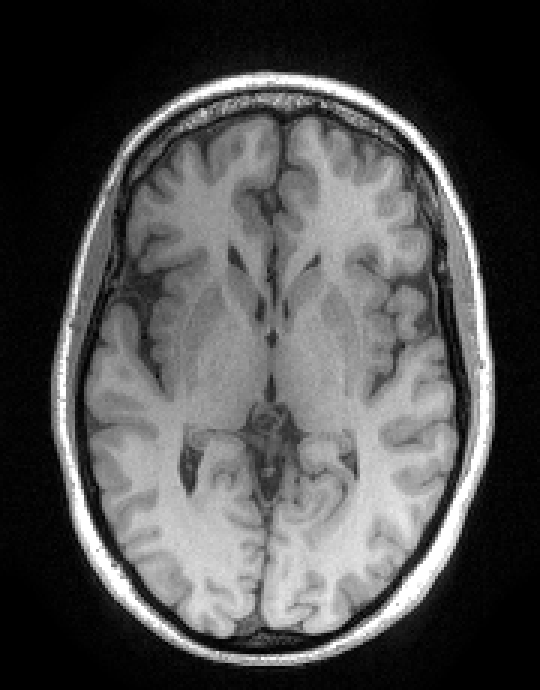}\includegraphics[width=.85in]{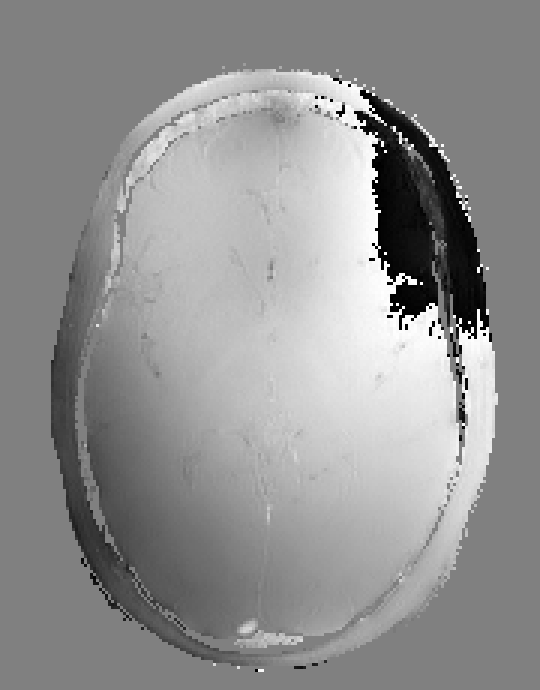}}
	\hfil
\subfloat[Noisy]{\includegraphics[width=.85in]{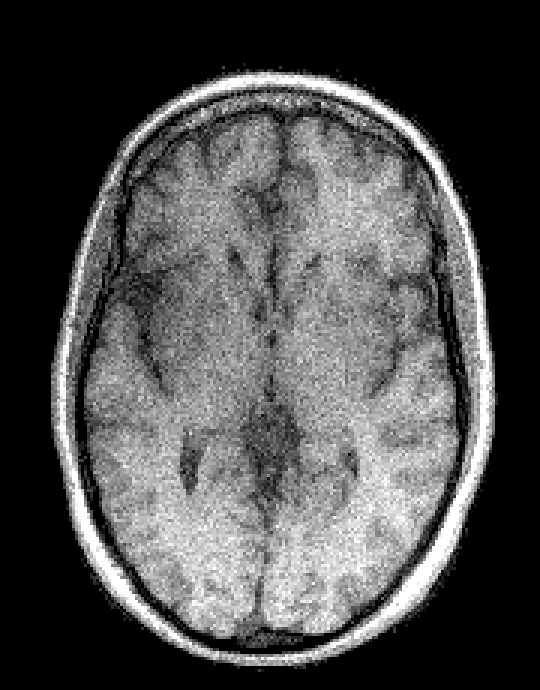}\includegraphics[width=.85in]{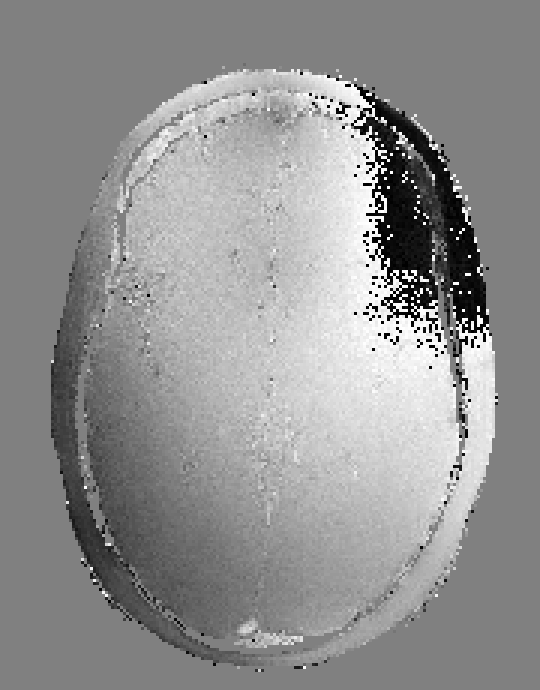}}
	\caption{The (a) ground truth and (b) noisy  magnitude and phase images used for the regularization-based MRI denoising scenario.}
	\label{fig:denoise_gold}
\end{figure}

\subsection{Regularization-based MRI Denoising}\label{sec:denoise}

In the second set of evaluations, we considered regularization-based denoising of a 230$\times$180 single-channel T1-weighted MR image, obtained by applying complex coil-combination to an 8-channel dataset and subsequently adding simulated complex Gaussian noise.  The ground truth and noisy images are shown in Fig.~\ref{fig:denoise_gold}. 

For reconstruction, the $\mathbf{A}$ matrix was an identity matrix.  Following Refs.~\cite{haldar2013,haldar2019,haldar2011,haldar2008}, the magnitude was regularized using a Huber-function penalty as given by Eq.~\eqref{eq:Huber_Mag_Tikhonov_Phs}, where a finite difference transformation was used to enforce spatial smoothness of the image.  Following Ref.~\cite{haldar2019,fessler2004,JustinDecovSLIM2011,FesslerSepMagPhs2012,ZibettiSepMagPhs2017}, the phase was regularized using a Tikhonov penalty as given by Eq.~\eqref{eq:Huber_Mag_Tikhonov_Phs2}, also using a finite difference transformation to enforce spatial smoothness.  All algorithms were initialized with the noisy image.

\begin{figure}[t]
	\centering
	\subfloat[]{\includegraphics[width=1.7in]{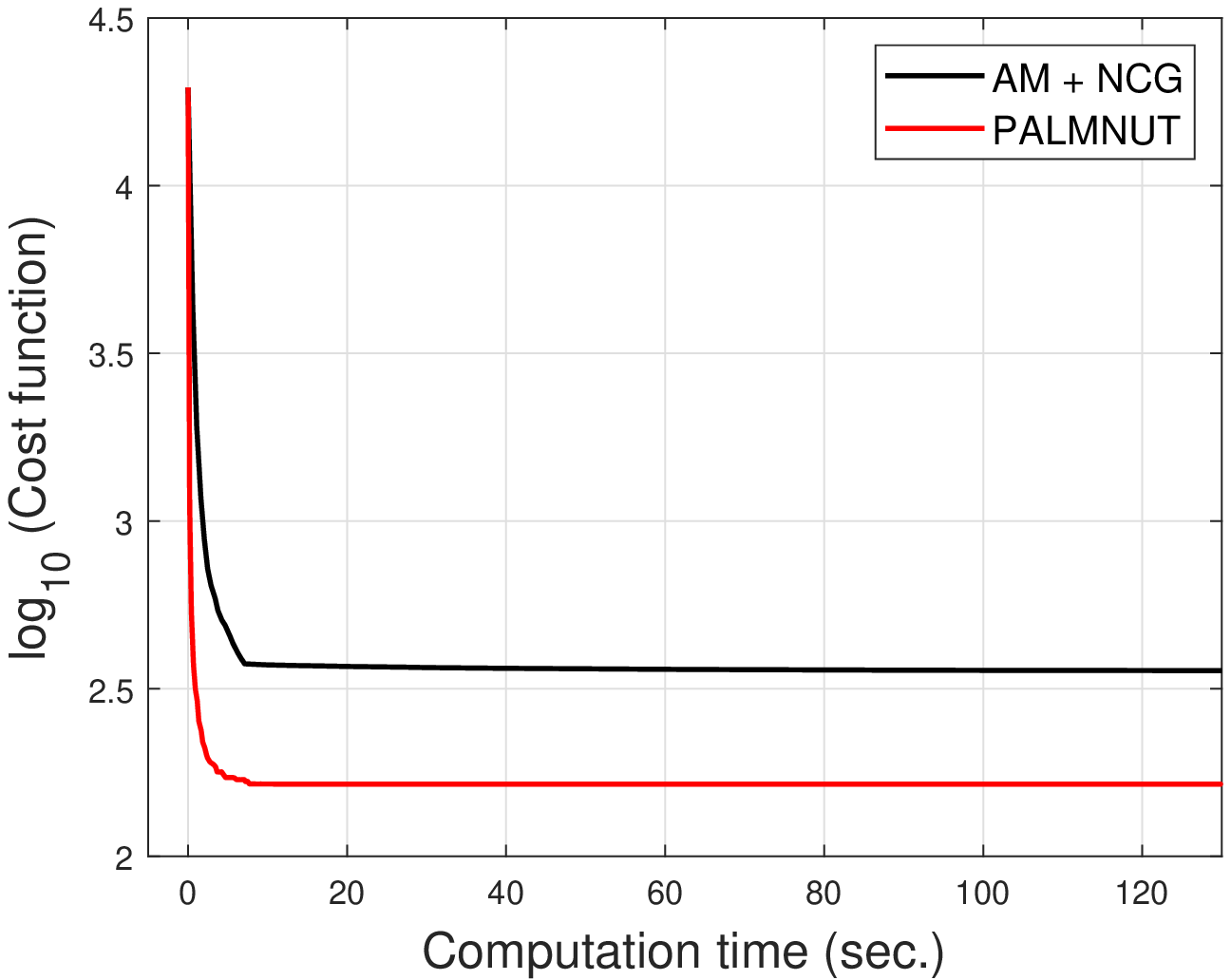}}
	\hfil
	\subfloat[]{\includegraphics[width=1.7in]{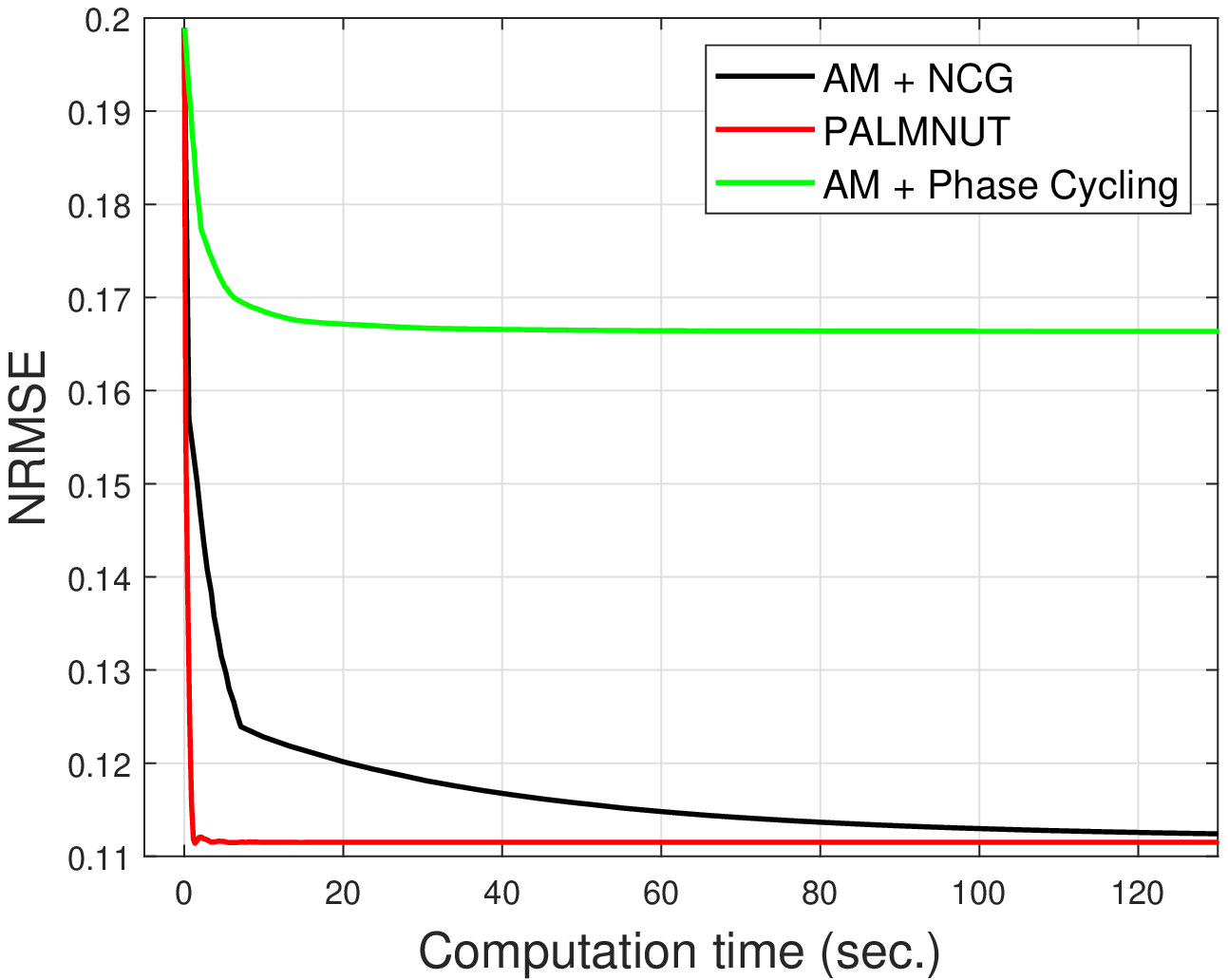}} \\
	\subfloat[PALMNUT]{\includegraphics[width=0.85in]{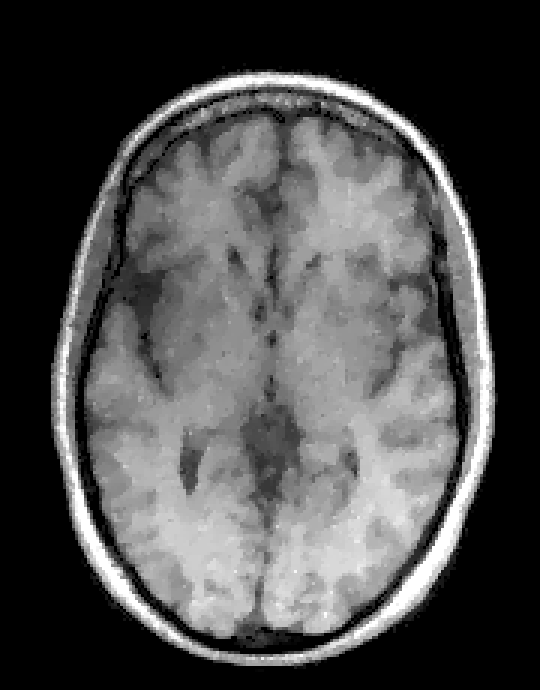}\includegraphics[width=0.85in]{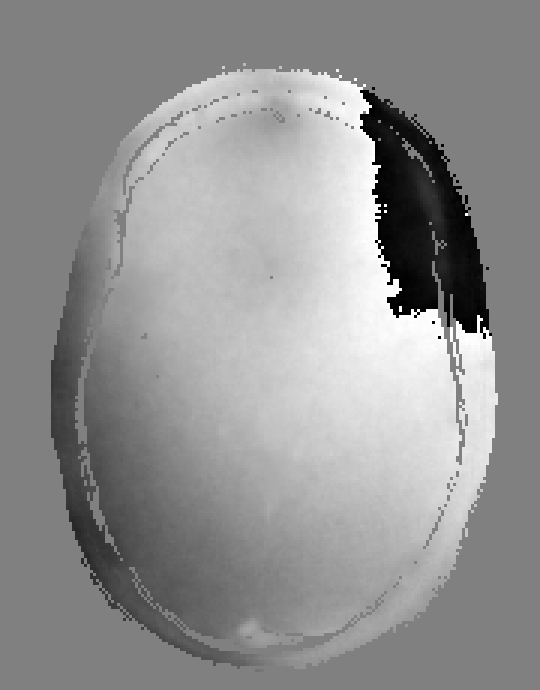}}
	\caption{Convergence plots and reconstructed images for the regularization-based MRI denoising scenario. The convergence plots show (a) the cost function value as a function of computation time and (b) the NRMSE value as a function of computation time.    Also shown are the (c) magnitude and phase images corresponding to  PALMNUT.}
	\label{fig:DenoiseConvCurv}
\end{figure}

Convergence results are shown in Fig.~\ref{fig:DenoiseConvCurv}, along with representative reconstruction results.  As can be seen, the results in this case are consistent with the previous case: PALMNUT and AM with NCG converged to similar NRMSE values, although PALMNUT was substantially faster, while AM with phase cycling was reasonably successful yet still substantially worse than the others in both speed and NRMSE.

\subsection{Phase-Corrected MR Image Combination}

\begin{figure}[t]
	\centering	
	\subfloat[Magnitude]{\includegraphics[width=1.6in]{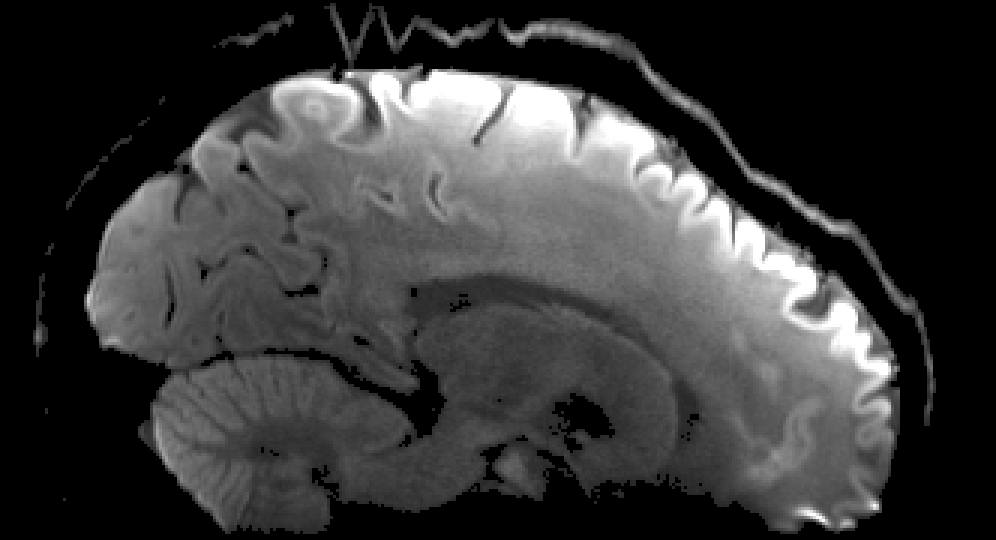}}
	\hfil
	\subfloat[Noisy]{\includegraphics[width=1.6in]{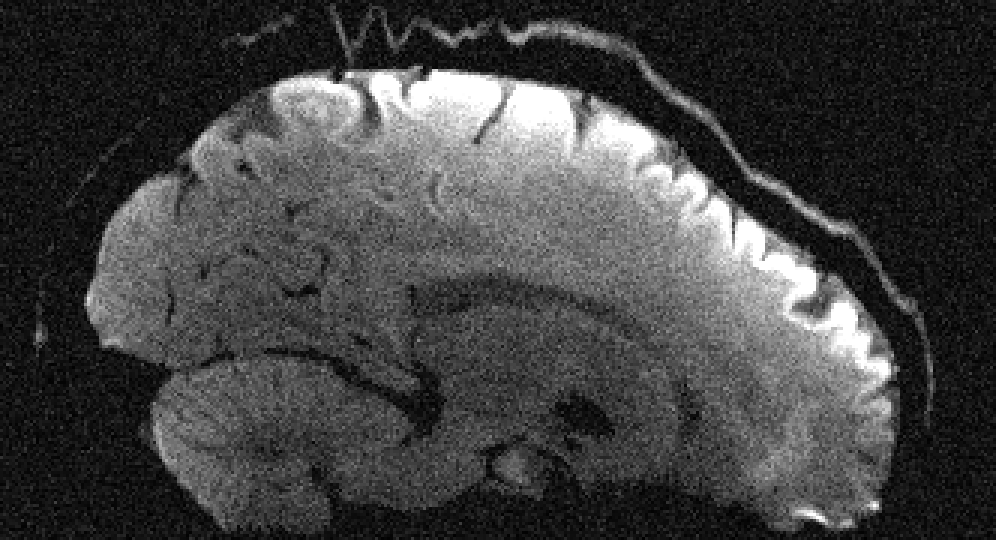}}\\
\subfloat[Phases]{\includegraphics[width=0.82in]{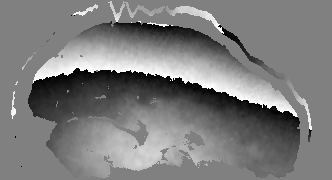}\hfil\includegraphics[width=0.82in]{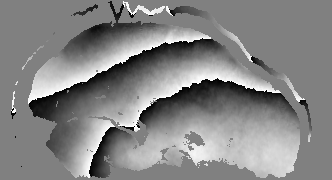}\hfil\includegraphics[width=0.82in]{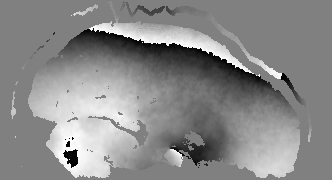}\hfil\includegraphics[width=0.82in]{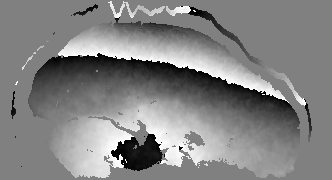}}
	\caption{(a) The  ground truth magnitude and (b) a representative noisy magnitude image, along with (c) the four different ground truth phase images used for the scenario with phase-corrected combination of multiple images. }\label{fig:combo_truth}
\end{figure}

In the third experiment, we simulated a scenario that is common in diffusion MRI, in which multiple measurements are made of the same image to enable averaging to improve SNR, but the phase varies randomly with each measurement due to experimental instabilities.  This case was simulated based on actual diffusion MRI magnitude and phase data from Ref.~\cite{haldar2019}.  We simulated a case with four repetitions, based on one ground truth magnitude image and four ground truth phase images, all with matrix size $180 \times 332$.  The magnitude image was combined with each of the different phase images to yield four different complex images, and then complex Gaussian noise was added to each result.  The ground truth images and a representative noisy image are shown in Fig.~\ref{fig:combo_truth}.

For reconstruction, we estimated a single shared magnitude image $\mathbf{m}$ and four different phase images $\mathbf{p}_j$ corresponding to the four different noisy measured images $\mathbf{b}_j$, $j=1,2,3,4$, by minimizing the cost function
\begin{equation}
R_1(\mathbf{m}) + \sum_{j=1}^4 \frac{1}{2} \|\mathbf{m}\odot e^{i \mathbf{p}_j} - \mathbf{b}_j \|_2^2 + R_2(e^{i \mathbf{p}_j}),
\end{equation}
where the magnitude and phase regularization penalties $R_1(\cdot)$ and $R_2(\cdot)$ were chosen in exactly the same way as for the denoising scenario from the previous subsection.  All optimization algorithms were initialized with the noisy phase images and by taking the average of the noisy magnitude images.

\begin{figure}
	\centering
	\subfloat[]{\includegraphics[width=1.7in]{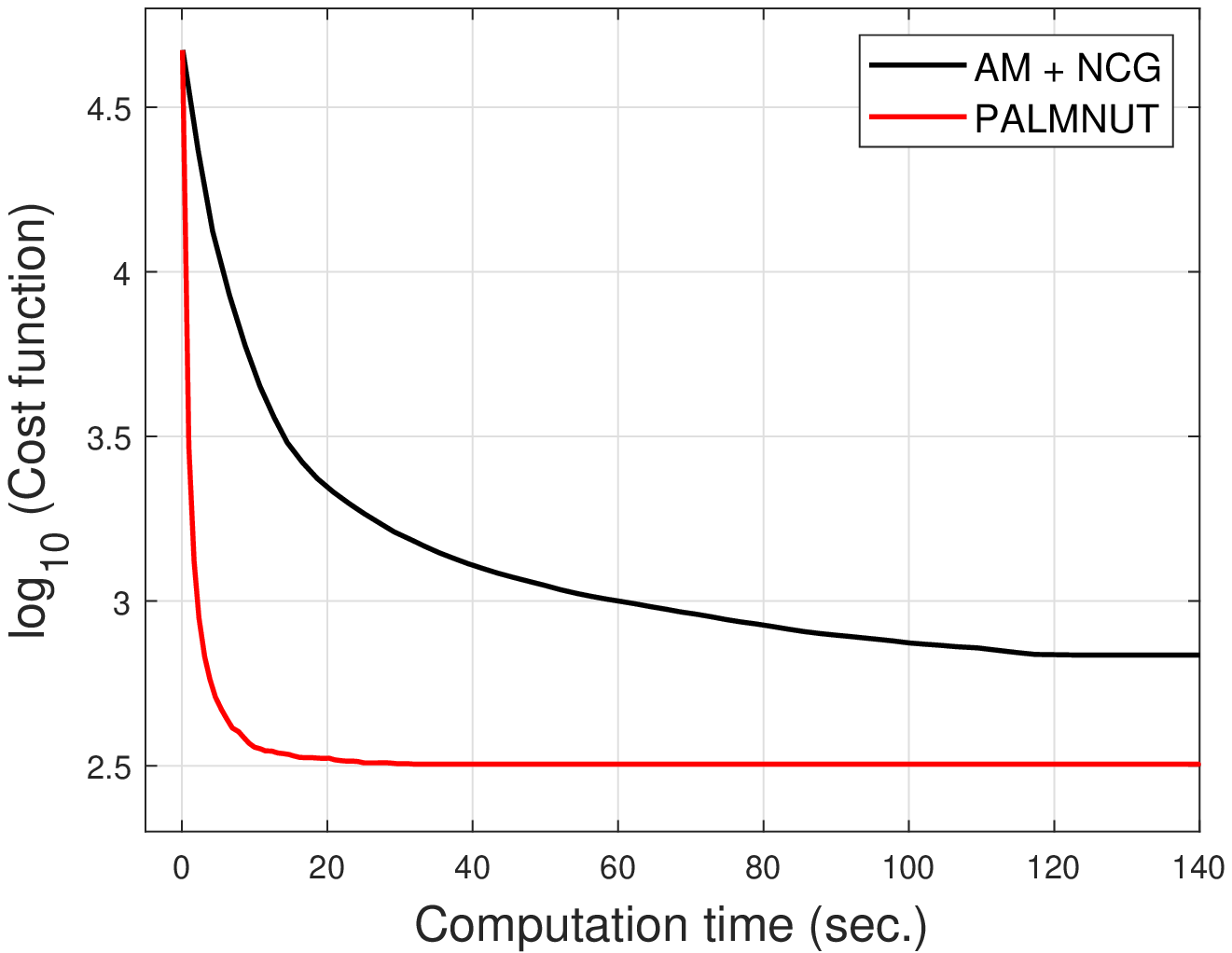}}
	\hfil 
	\subfloat[]{\includegraphics[width=1.7in]{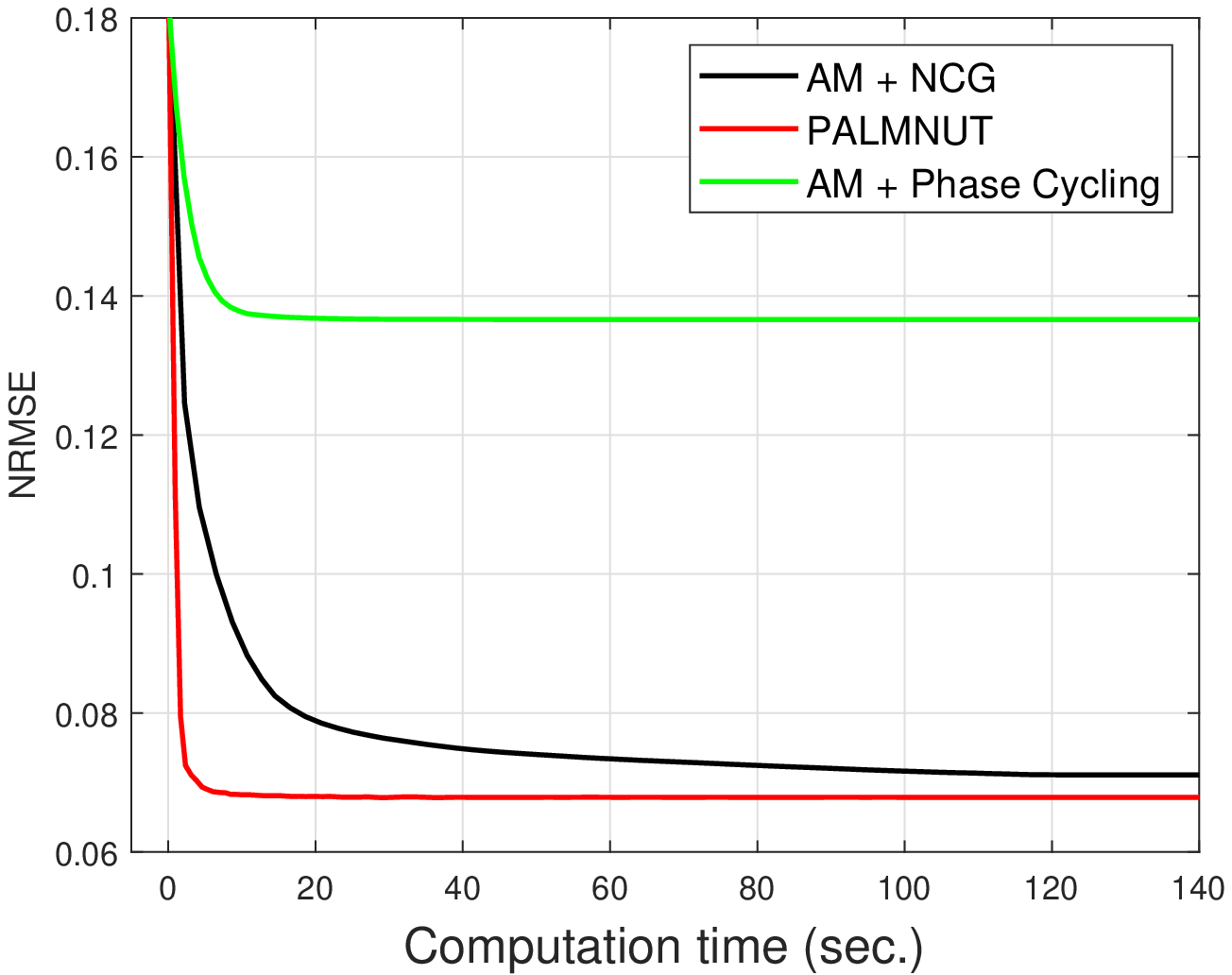}}\\
	\subfloat[PALMNUT]{\includegraphics[width=1.7in]{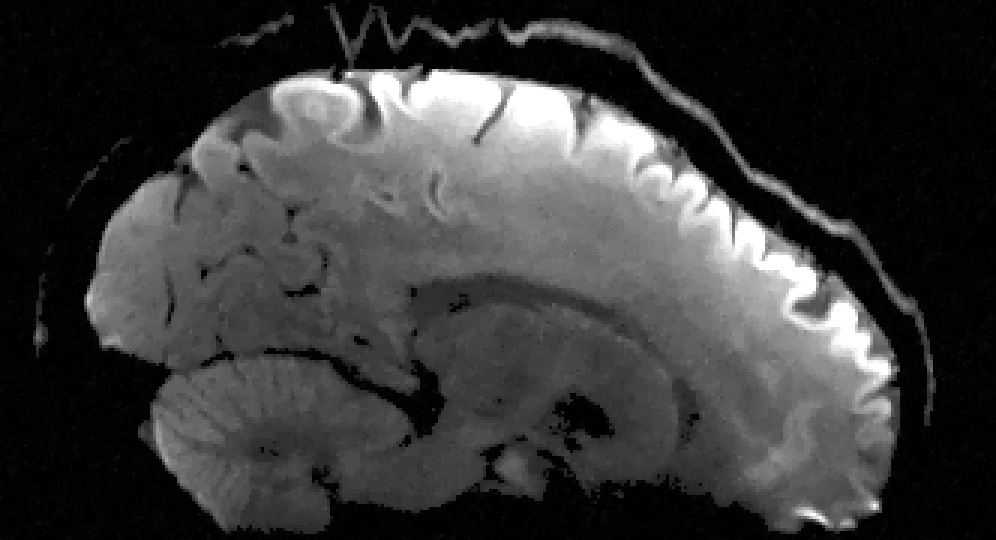}}
	\caption{Convergence plots and reconstructed images for the phase-corrected MRI image combination scenario.  The convergence plots show (a) the cost function value as a function of computation time and (b) the NRMSE value as a function of computation time.  Also shown is (c) the combined magnitude image obtained from PALMNUT.}
	\label{fig:ShareMagConvCurv}
\end{figure}

The convergence plots shown in Fig.~\ref{fig:ShareMagConvCurv} show similar characteristics to those observed in the previous scenarios, with PALMNUT having a distinct advantage over the two alternative algorithms.

\section{Discussion}

\begin{figure}[t]
	\centering
	\subfloat[Undersampled MRI Reconstruction]{\includegraphics[width=1.7in]{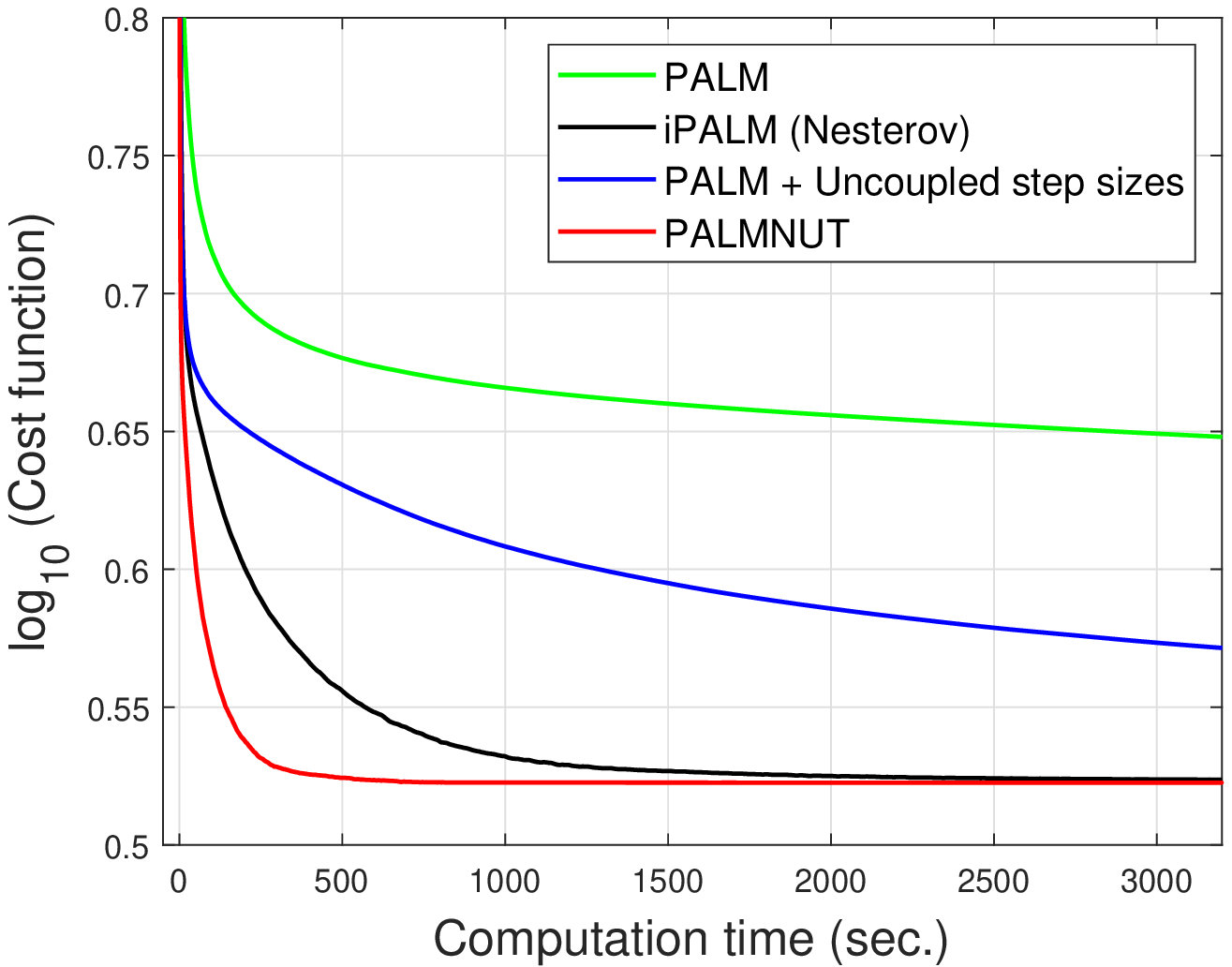}\includegraphics[width=1.7in]{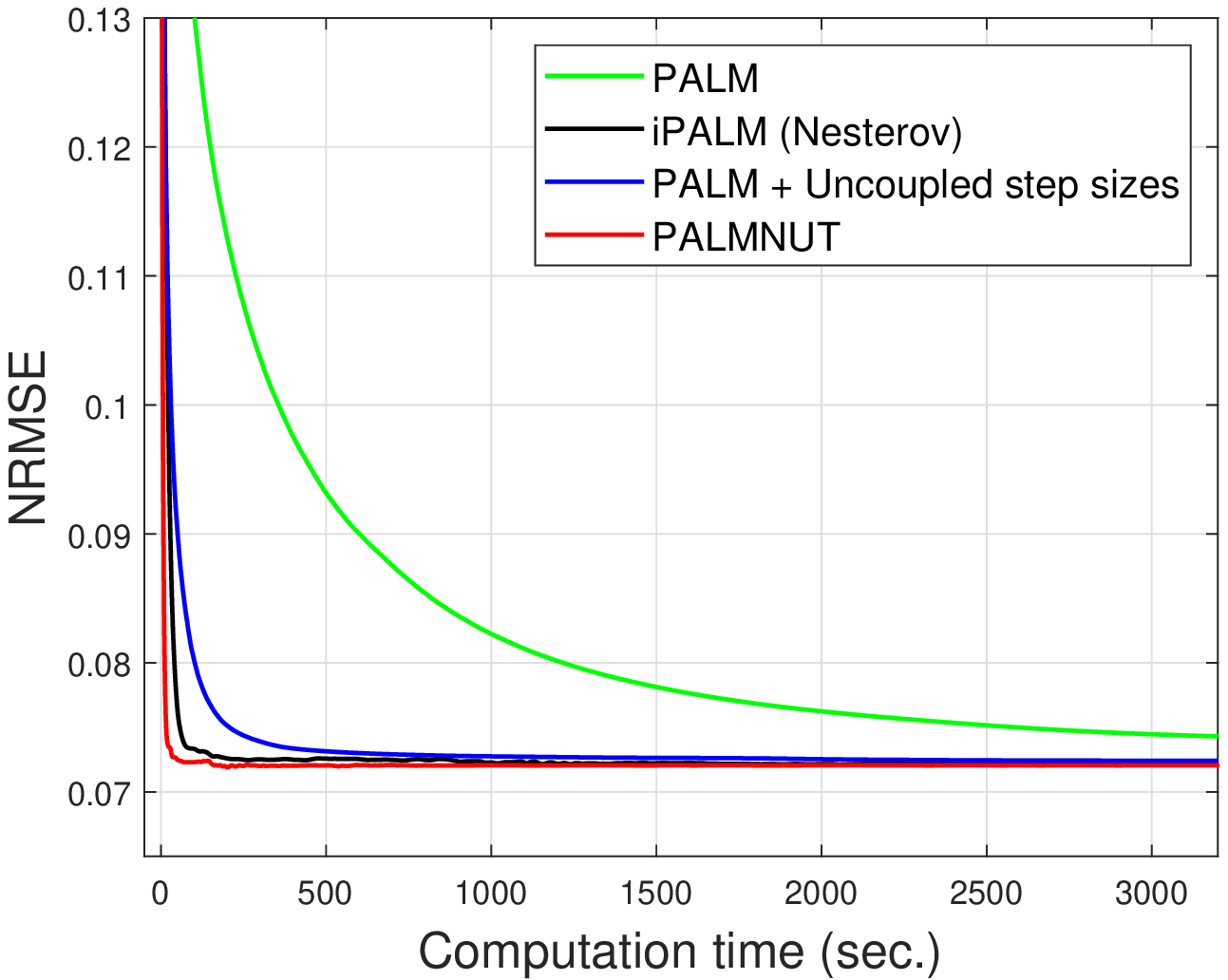}} \\
	\subfloat[Regularization-based MRI Denoising]{\includegraphics[width=1.7in]{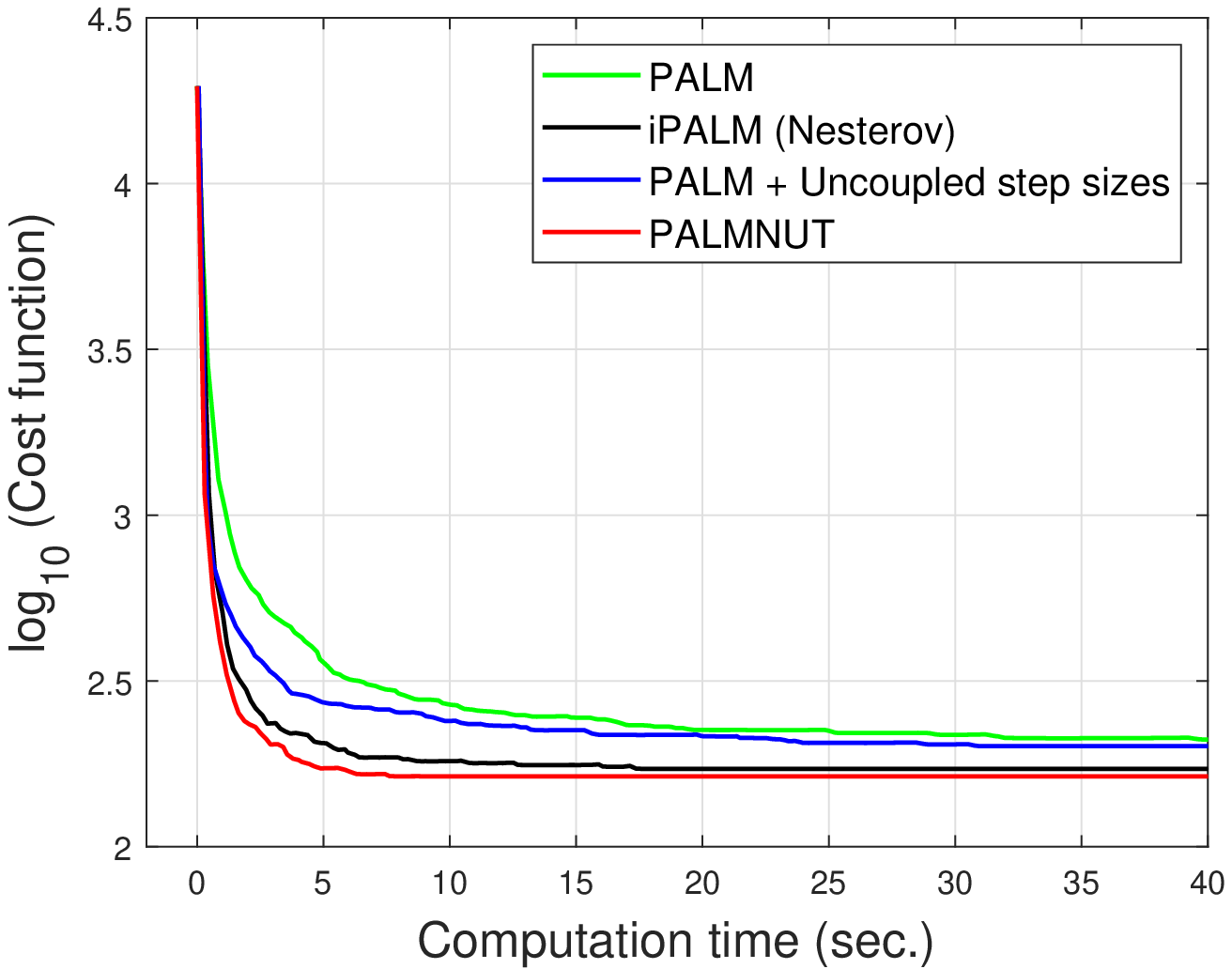}\includegraphics[width=1.7in]{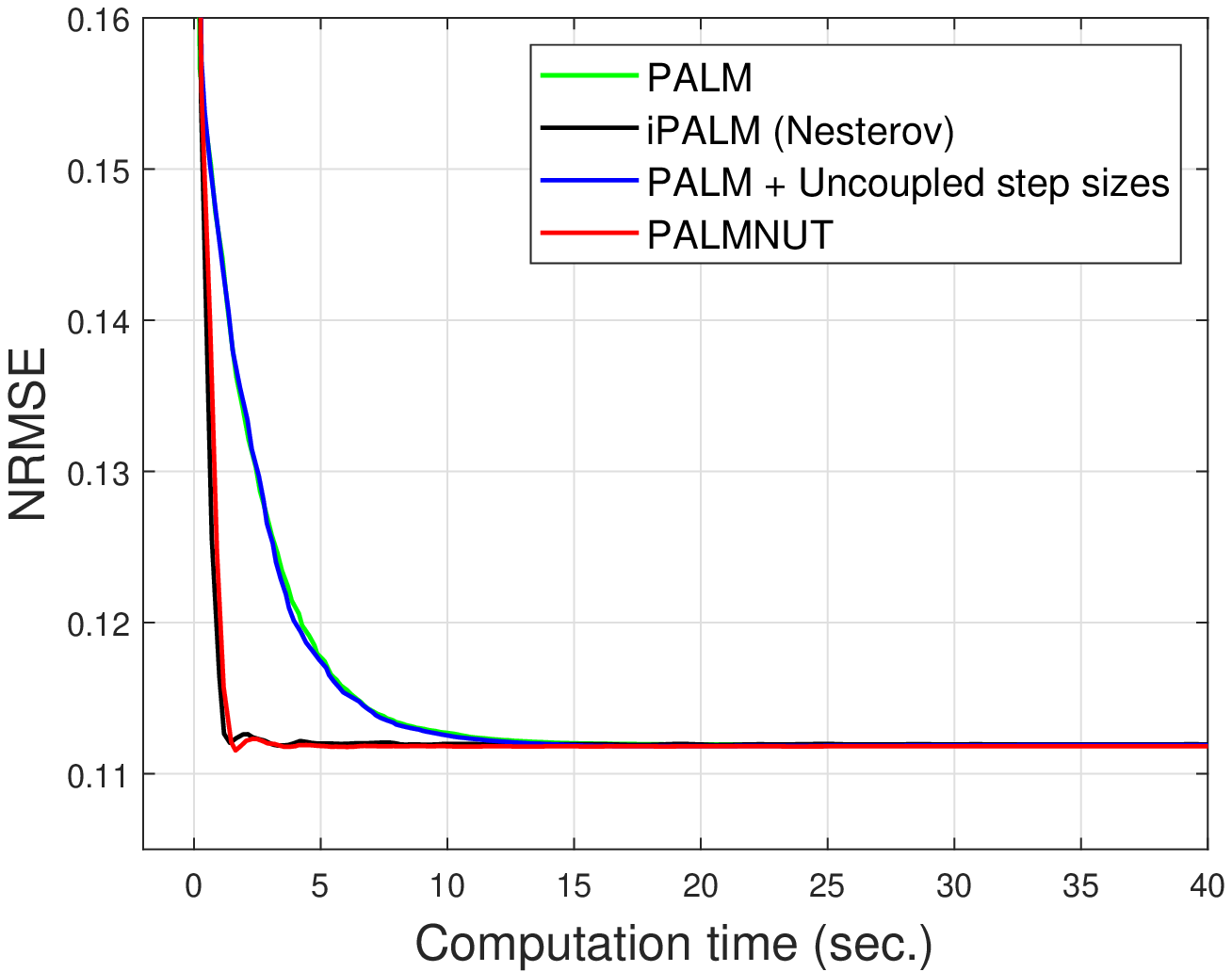}}
	\caption{The convergence plots show the cost function value and the NRMSE value as a function of computation time for (a) the undersampled MRI reconstruction scenario and (b) the regularization-based MRI denoising scenario.}
	\label{fig:ablation}
\end{figure}

The results of the previous section demonstrated that PALMNUT can have major advantages relative to AM methods.  However, PALMNUT represents a combination of three different ideas (PALM, Nesterov's momentum, and uncoupled stepsizes), and the previous experiments did not investigate which parts of PALMNUT contribute most to its performance.  In order to gain more insight, we did another set of experiments in which we compared PALMNUT against the original PALM algorithm (Alg.~\ref{alg:PALM}), PALM with Nesterov's momentum but without uncoupled stepsizes (iPALM), and PALM with uncoupled stepsizes but without Nesterov's momentum (Alg.~\ref{alg:PALMUT}).  Results are shown in Fig.~\ref{fig:ablation} for both the undersampled MRI reconstruction (Section~\ref{sec:under}) and the regularization-based MRI denoising (Section~\ref{sec:denoise}) scenarios described previously.  From these plots, we observe that both uncoupled step sizes and Nesterov's momentum independently improve the convergence speed relative to the original PALM method, with Nesterov's momentum contributing a little more than the use of uncoupled step sizes.  However, PALMNUT's combination of all of these elements leads to the best overall results.

An interesting phenomenon we observed with PALMNUT is that the magnitude estimate frequently converged faster than the phase estimate (results not shown).  As a result, it could be potentially beneficial to update the phase estimate more frequently than the magnitude estimate, although we believe that such an exploration is beyond the scope of this paper.   

One of the key ingredients of PALMNUT is the use of coordinatewise step sizes based on coordinatewise Lipschitz-like bounds in the form of Eqs.~\eqref{eq:Lipd1} and \eqref{eq:Lipd2}.  Although these bounds were formulated in the context of PALM, we believe that Lemma 1 (found in the Appendix and used in the proof of Theorem 1) represents a novel coordinatewise majorization relationship, which could also potentially be useful to construct majorants in more general optimization scenarios.

Finally, we should mention that while PALM combined with uncoupled step sizes has guaranteed monotonic convergence properties, the convergence of the PALMNUT approach (which additionally includes Nesterov's momentum) has not yet been theoretically proven.  This represents another potentially interesting topic for future research.

\section{Conclusion}

We proposed and evaluated a new algorithm called PALMNUT, which combines the PALM algorithm with Nesterov's momentum with uncoupled coordinatewise step sizes derived from coordinatewise Lipschitz-like bounds.  Although our approach is general and can be applied to other computational imaging scenarios, our evaluation studies focused on MRI scenarios involving separate regularization of the image magnitude and phase.  Our empirical results demonstrated that PALMNUT consistently had substantial advantages over previous approaches based on alternating minimization across several different MRI scenarios. As a result, we expect that PALMNUT will be useful for these kinds of MRI scenarios, and may also prove useful for more general computational imaging problems with similar optimization structure.

\onecolumn
\appendices
\section{Proof of Thm.~\ref{thm:UT_Conv}}\label{app:proof}
Theorem~\ref{thm:UT_Conv} is an immediate consequence of the following Lemma. 
\begin{lemma} \label{Lema:DesctIneq}
Let $Q(\mathbf{x}): \mathbb{R}^N \rightarrow \mathbb{R}$ be smooth, and assume that a non-negative vector $\mathbf{L} \in \mathbb{R}^{N}$ exists such that
\begin{equation}
\begin{split}
\inner{\nabla Q(\mathbf{x}_1) - \nabla Q(\mathbf{x}_2), \mathbf{x}_1 - \mathbf{x}_2} \leq \norm{\sqrt{\mathbf{L}} \odot (\mathbf{x}_1 - \mathbf{x}_2)}_2^2
\end{split}
\end{equation}
for $\forall \mathbf{x}_1,\mathbf{x}_2 \in \mathbb{R}^N$.  Then for $\forall \mathbf{x},\hat{\mathbf{x}}_k \in \mathbb{R}^N$ and $\forall \mathbf{c} \geq \mathbf{L}$ (elementwise), we have
\begin{equation} \label{eq:CompDesct}
\begin{split}
Q(\mathbf{x}) &\leq Q(\hat{\mathbf{x}}_k) + \inner{\mathbf{x}-\hat{\mathbf{x}}_k,\nabla Q(\hat{\mathbf{x}}_k)} + \frac{1}{2} \norm{\sqrt{\mathbf{c}}\odot\left( \mathbf{x}-\hat{\mathbf{x}}_k\right)}_{2}^{2}.
\end{split}
\end{equation}
Further, the right hand side of Eq.~\eqref{eq:CompDesct} can be written compactly as
\begin{equation}\label{eq:CompDesct2}
\begin{split}\\
Q(\hat{\mathbf{x}}_k) + \inner{\mathbf{x}-\hat{\mathbf{x}}_k,\nabla Q(\hat{\mathbf{x}}_k)} + \frac{1}{2} \norm{\sqrt{\mathbf{c}}\odot\left( \mathbf{x}-\hat{\mathbf{x}}_k\right)}_{2}^{2}
&= \tau_k + \frac{1}{2}\norm{\sqrt{\mathbf{c}} \odot (\mathbf{x} - \mathbf{w}_k)}_2^2,
\end{split}
\end{equation}
where 
\begin{equation}
\mathbf{w}_k = \hat{\mathbf{x}}_k - \mathrm{diag}(\mathbf{c})^{-1} \nabla Q(\hat{\mathbf{x}}_k)
\end{equation}
and $\tau_k$ is a constant that does not depend on the variable $\mathbf{x}$,  given by
\begin{equation}
\tau_k = Q(\hat{\mathbf{x}}_k) - \inner{\hat{\mathbf{x}}_k,\nabla Q(\hat{\mathbf{x}}_k)}+\frac{1}{2} \norm{\sqrt{\mathbf{c}}\odot \hat{\mathbf{x}}_k}_2^2 - \frac{1}{2}\norm{\sqrt{\mathbf{c}} \odot \mathbf{w}_k}_2^2.
\end{equation}

\end{lemma}

\begin{proof}
	Inspired by the proof of Proposition A.24 from \cite{BertsekasNonlinearProgram}, define $\alpha(t) \triangleq Q(t\mathbf{x} + (1-t)\hat{\mathbf{x}}_k)$. Then $\alpha(0) = Q(\hat{\mathbf{x}}_k)$, $\alpha(1) = Q(\mathbf{x})$, and 
	\begin{equation}
	 \frac{d}{dt}\alpha(t)  = \inner{\nabla Q(t\mathbf{x} + (1-t) \hat{\mathbf{x}}_k), \mathbf{x} - \hat{\mathbf{x}}_k}.
	\end{equation}

	We have that
	\begin{equation}
	\begin{split}
	 Q(\mathbf{x}) - Q(\hat{\mathbf{x}}_k) &= \int_{0}^1 \frac{d}{dt}\alpha(t) dt \\
	 & = \int_0^1 \inner{\nabla Q(t\mathbf{x} + (1-t) \hat{\mathbf{x}}_k), \mathbf{x} - \hat{\mathbf{x}}_k} dt \\
	 & = \int_0^1 \inner{\nabla Q(t\mathbf{x} + (1-t) \hat{\mathbf{x}}_k) - \nabla Q(\hat{\mathbf{x}}_k), \mathbf{x} - \hat{\mathbf{x}}_k} dt + \inner{\nabla Q(\hat{\mathbf{x}}_k),\mathbf{x}-\hat{\mathbf{x}}_k} \\
	 & = \int_0^1 \frac{1}{t} \inner{\nabla Q(t\mathbf{x} + (1-t) \hat{\mathbf{x}}_k) - \nabla Q(\hat{\mathbf{x}}_k), t \mathbf{x} - t\hat{\mathbf{x}}_k} dt + \inner{\mathbf{x}-\hat{\mathbf{x}}_k,\nabla Q(\hat{\mathbf{x}}_k)}\\
	 & =  \int_0^1 \frac{1}{t}  \inner{\nabla Q(t\mathbf{x} + (1-t) \hat{\mathbf{x}}_k) - \nabla Q(\hat{\mathbf{x}}_k), t \mathbf{x} + (1- t)\hat{\mathbf{x}}_k - \hat{\mathbf{x}}_k} dt +\inner{\mathbf{x}-\hat{\mathbf{x}}_k,\nabla Q(\hat{\mathbf{x}}_k)} \\
	 & \leq  \int_{0}^1 \frac{1}{t} \norm{\sqrt{\mathbf{L}} \odot (t \mathbf{x} + (1- t)\hat{\mathbf{x}}_k - \hat{\mathbf{x}}_k)}_2^2 dt+ \inner{\mathbf{x}-\hat{\mathbf{x}}_k,\nabla Q(\hat{\mathbf{x}}_k)} \\	 
	 & =   \int_{0}^1 t \norm{\sqrt{\mathbf{L}} \odot (\mathbf{x}  - \hat{\mathbf{x}}_k)}_2^2 dt+ \inner{\mathbf{x}-\hat{\mathbf{x}}_k,\nabla Q(\hat{\mathbf{x}}_k)} \\
	 & = \frac{1}{2}\norm{\sqrt{\mathbf{L}} \odot (\mathbf{x}  - \hat{\mathbf{x}}_k)}_2^2 + \inner{\mathbf{x}-\hat{\mathbf{x}}_k,\nabla Q(\hat{\mathbf{x}}_k)} \\
	 & \leq \frac{1}{2}\norm{\sqrt{\mathbf{c}} \odot (\mathbf{x}  - \hat{\mathbf{x}}_k)}_2^2 + \inner{\mathbf{x}-\hat{\mathbf{x}}_k,\nabla Q(\hat{\mathbf{x}}_k)}.
	 \end{split}
	\end{equation}
	This derivation proves  Eq.~\eqref{eq:CompDesct}, while the simplifications leading to Eq.~\eqref{eq:CompDesct2} come simply from completing the square.
	\end{proof}
	
\twocolumn 


\end{document}